\documentclass[fleqn,10pt]{wlscirep}
\usepackage[utf8]{inputenc}
\usepackage[T1]{fontenc}

\pdfoutput=1

\usepackage{graphicx,graphics}
\usepackage{bm,enumerate,amsmath,amssymb,amsthm}
\usepackage{epsfig}

\usepackage{CJKutf8}
\usepackage[utf8]{inputenc}
\usepackage{graphicx}
\usepackage{graphics}
\usepackage{tabularx}
\usepackage{bm}
\usepackage{nameref}
\usepackage{caption}
\usepackage{xr}

\usepackage{booktabs}
\usepackage{multirow}

\usepackage{adjustbox}

\usepackage{here}
\usepackage{float}

\usepackage{color, soul}
\usepackage{gensymb}

\usepackage{hyperref}

\usepackage{amsmath}
\usepackage{amssymb}

\usepackage{adjustbox}

\usepackage{color, soul}
\usepackage{gensymb}

\usepackage{booktabs}

\graphicspath{{./}}

\usepackage[superscript,biblabel]{cite}

\newcommand{\beginsupplement}{%
        \setcounter{table}{0}
        \renewcommand{\thetable}{S\arabic{table}}%
        \setcounter{figure}{0}
        \renewcommand{\thefigure}{S\arabic{figure}}%
     }

\title{Accurate predictive model of band gap with selected important features based on explainable machine learning}

\author[1,*]{Joohwi Lee}
\author[1]{Kaito Miyamoto}

\affil[1]{Toyota Central R\&D Labs., Inc., Yokomichi 41--1, Nagakute, Aichi, 480--1192, Japan}
\affil[*]{email: j-lee@mosk.tytlabs.co.jp}


\begin{abstract}
In the rapidly advancing field of materials informatics, nonlinear machine learning models have demonstrated exceptional predictive capabilities for material properties. 
However, their black-box nature limits interpretability, and they may incorporate features that do not contribute to---or even deteriorate---model performance.
This study employs explainable ML (XML) techniques, including permutation feature importance and the SHapley Additive exPlanation, applied to a pristine support vector regression model designed to predict band gaps at the GW level using 18 input features.
Guided by XML-derived individual feature importance, a simple framework is proposed to construct reduced-feature predictive models. 
Model evaluations indicate that an XML-guided compact model, consisting of the top five features, achieves comparable accuracy to the pristine model on in-domain datasets (0.254 vs. 0.247 eV) while showing improved generalization with lower prediction errors on out-of-domain data (0.348 vs. 0.460 eV). 
Additionally, the study underscores the necessity for eliminating strongly correlated features (correlation coefficient greater than 0.8) to prevent misinterpretation and overestimation of feature importance before applying XML.
This study highlights XML’s effectiveness in developing simplified yet highly accurate machine learning models by clarifying feature roles, thereby reducing computational costs for feature acquisition and enhancing model trustworthiness for materials discovery.
\end{abstract}
\begin{document}

\flushbottom
\maketitle

\thispagestyle{empty}
    
\section*{\label{sec_intro} Introduction}
The rapid advancement of materials informatics has been driven by machine learning (ML), enabling significant improvements in predicting the properties of materials and accelerating their discovery. 
Nonlinear ML models, such as support vector machines\cite{svr1,svr2,svr3} and neural networks,\cite{deep} offer superior predictive performance compared to linear models, and have thus gained popularity. 
However, the black-box nature of such models limits their interpretability and explainability.\cite{xai1-book,xai2,xai3}

To address this challenge, explainable ML (XML), also known as explainable artificial intelligence (XAI), has gained attention. 
XML enhances the transparency of the model and fosters scientific understanding by revealing relationships among material structures, compositions, and properties. 
The use of XML is motivated by the following factors:
(i) Understanding the decision-making mechanisms of ML models is essential for building reliable predictive models. 
(ii) Interpretability aids in debugging unexpected predictions and abnormalities. 
(iii) XML enables the identification of key predictive features through feature importance analysis, facilitating the reduction of feature preparation costs by eliminating unnecessary features.
Notably, XML methods enable nonlinear ML models to explain themselves in an interpretable manner through feature importance metrics, analogous to how linear models use coefficient values to indicate feature significance.

The band gap ($E_\textrm{g}$) is a critical property that determines the electrical conductivity of a material and its suitability for electronic and optoelectronic applications.
Although density functional theory (DFT)\cite{kohn-dft} provides a straightforward approach for estimating the band gap of materials, such predictions are often significantly underestimated. 
Methods employing hybrid functionals\cite{HSE03} and GW-level\cite{GW} calculations offer improved accuracy, albeit at a high computational cost.
In particular, the band gap predicted using GW-level calculations have much higher fidelity with the experimental values compared to those predicted using hybrid functionals, but the computational cost of the former is substantially greater, severely limiting the accessible size of the dataset.
Consequently, ML models have become indispensable for the efficient prediction of accurate band gap values in materials discovery. 
Lee $et$ $al$.\cite{lee2016GWgap} developed ML models, including support vector regression (SVR)\cite{svr1,svr2,svr3}, for predicting band gap at the G$_0$W$_0$-level ($E_\textrm{g}^{\textrm{GW}}$).\cite{GW} 
Their approach incorporated compound-level features derived from DFT calculations alongside elemental features based on the means and standard deviations of constituent element properties. 
When utilizing only the band gap computed via generalized gradient approximation using the Perdew–Burke–Ernzerhof (PBE) exchange-correlation functional ($E_\textrm{g}^{\textrm{PBE}}$)\cite{PBE} as a single feature, the model achieved a root mean square error (RMSE) of 0.60 eV on the test dataset. 
By integrating 17 additional features, the accuracy of the optimized model was improved, where the RMSE was reduced to 0.24 eV. 
However, the contributions of these 17 features to the predictive performance of the model remain insufficiently explored. 
Some features may have a negligible impact because their roles overlap owing to strong correlations among the features, or because they do not substantially contribute to accurately predicting the intended prediction objective.
Reducing the number of input features could decrease the computational cost of feature preparation and improve the model generalization by mitigating overfitting, which is particularly important for accurate prediction of out-of-domain (OOD) data.
Therefore, enhancing the interpretability of such ML models is essential for gaining deeper insights into materials design.
Identifying the key predictive features and simplifying the model can reduce the complexity while maintaining high accuracy.

XML methods have been increasingly applied to band gap predictive models in recent studies. 
Obada $et$ $al$.\cite{Eg_obada} employed XML in the analysis of models for predicting band gap for 199 $ABX_3$ perovskites using element-specific features.
However, their analysis was confined to the perovskite family, limiting generalization to broader chemistries.
Choubisa  $et$ $al$.\cite{Eg_choubisa} introduced the Deep Adaptive Regressive Weighted Intelligent Network, integrating an ML surrogate model, an evolutionary algorithm-based search method, and an interpretable design rule for materials with direct band gap. 
To enhance the interpretability of the model, they employed Spearman’s rank correlation coefficient ($r_s$) and permutation feature importance (PFI).\cite{pfi}
However, their applications also largely focused on perovskite-like semiconductors, again restricting the material domain.
Zhang $et$ $al$.\cite{Eg_zhang} proposed an interpretable $\Delta$-ML model that correlates the band gap values obtained with the Heyd–Scuseria–Ernzerhof (HSE) hybrid functional~\cite{HSE03} and the PBE functional~\cite{PBE} for two-dimensional materials, utilizing the sure independence screening and sparsifying operator (SISSO) algorithm.~\cite{SISSO} 
Shi $et$ $al$.~\cite{Eg_shi} applied XML techniques, including SHapley Additive exPlanations (SHAP)~\cite{shap1} and SISSO, to models for predicting the thermodynamic stability and band gap computed using the HSE functional for Janus III--IV van der Waals heterostructures.
Both of these studies demonstrate the utility of XML in 2D contexts. 
Nevertheless, the scope of both studies is confined to layered materials rather than bulk inorganic compounds.
The abovementioned studies demonstrate the value of XML but are generally restricted to narrow material classes.
In addition, these studies did not cross-check the feature importance using multiple XML methods, for example, by combining PFI and SHAP.

Jihad $et$ $al$.\cite{jihad2024dft} also proposed a compact five-feature model for correcting $E_\textrm{g}^{\textrm{PBE}}$ to $E_\textrm{g}^{\textrm{GW}}$ from the same compound dataset used herein, but their feature selection was based on physics-guided reasoning and manual correlation control rather than XML-based ranking.
Therefore, an XML framework applicable to more generalized and diverse datasets is clearly needed.

In this study, we address this need by applying XML to a nonlinear SVR model for predicting $E_\textrm{g}^{\textrm{GW}}$ for OOD inorganic materials as well as in-domain systems, including binary and ternary compounds.
SVR\cite{svr1,svr2,svr3} is chosen after benchmarking against interpretable linear models such as the ordinary least-square and least absolute shrinkage and selection operator (LASSO)\cite{lasso} regressions, as the former consistently achieved higher accuracy and continued to improve with larger training data.\cite{lee2016GWgap}
The strong predictive ability of the selected model makes it a particularly suitable baseline for demonstrating the value of XML, despite its lack of interpretability.
The XML methods, PFI\cite{pfi} and SHAP\cite{shap1}, are combined in a systematic manner to guide feature ranking and selection, and the consistency of the feature importance rankings is compared by cross-checking the two methods to ensure reliability.
Important features are then identified based on these XML analyses, retaining only those with consistently high importance across both metrics.
The derived feature importances are also cross-checked against the coefficient magnitudes obtained from the interpretable LASSO regression.
Because highly correlated features can distort importance estimation using XML, such features are eliminated prior to the XML analysis.
The effect of this approach is explicitly demonstrated in practical examples in the present study.
Through statistical validation and subsampling, the predictive performance and generalization ability of the reduced-feature models are quantitatively evaluated against those of the pristine model using OOD data. 
Taken together, these steps form a simple and explicit framework for constructing compact yet interpretable predictive models.

\section*{\label{sec_methods} Methods}

\subsection*{\label{subsec_modeling}Band gap predictive model}

This study revisits the predictive model proposed by Lee $et$ $al$.\cite{lee2016GWgap} for estimating $E_\textrm{g}^\textrm{GW}$ for 270 binary and ternary inorganic compounds. 
The pristine model employed SVR\cite{svr1,svr2,svr3} with 18 input features ($x_j$).\cite{lee2016GWgap2}
Fourteen features are derived from the mean ($\bar{x}$) and standard deviation ($\sigma(x)$) of the elemental properties, including the absolute oxidation number ($\mid$$n$$\mid$), atomic number ($Z$), period in the periodic table ($p$), atomic mass ($m$), van der Waals radius ($r$), electronegativity ($\chi$), and ionization energy ($I$). 
Notably, these features do not require DFT calculations.
Compound-specific properties obtained from DFT calculations are also incorporated, including $E_\textrm{g}^\textrm{PBE}$, the bandgap computed using the modified Becke--Johnson exchange-correlation functional ($E_\textrm{g}^\textrm{mBJ}$),\cite{mBJ} volume per atom ($V$), and cohesive energy ($E_\textrm{coh}$). 

The dataset is split into 75\% for training and 25\% for testing.
Predictions are conducted over 20 iterations, each using an independently randomized training/test split.
To ensure consistency across runs, the random seeds are set to 0--19.
For the SVR, a radial basis function kernel is employed.
The hyperparameters are optimized via grid search with ten-fold cross-validation on the training set over the following ranges: $C$ $\in$ $\{10^{-5}, 10^{-4}, \dots, 10^{4}\}$, $\gamma$ $\in$ $\{10^{-6}, 10^{-5}, \dots, 10^{5}\}$, and $\epsilon$ $\in$ $\{0.001, 0.005, 0.01\}$.
For each iteration, the optimal hyperparameters are determined by ten-fold cross-validation on the training set and subsequently used to train the final SVR model on the full training set.
Before introduction into the model, all features are standardized via $Z$--score normalization using the StandardScaler implemented in the Python Scikit-learn library.\cite{scikit-learn}
The entire process is implemented in the same library.
The exact data splits and the corresponding optimal hyperparameters are provided in the publicly available code repository.

In this study, the feature dimensionality of the pristine model is reduced through a two-step process. 
First, all feature pairs are subjected to correlation analysis and one feature is removed from each pair exhibiting strong correlation.
Second, feature importance is assessed using PFI and SHAP, retaining only the most significant features. 
The impact of this reduced-feature set on the predictive performance of the model is then evaluated. 

For comparison, LASSO regression, which can perform both shrinkage and embedded feature selection based on $L_1$ penalty, is also employed as an interpretable linear baseline. 
This method is trained using the same standardized features and 20 randomized data splits as used for SVR. 
The regularization strength $\alpha$ is optimized by 10-fold cross-validation using 200 logarithmically spaced values between $10^{-6}$ and $10^{2}$, and the maximum iteration number is set to 100{,}000 to ensure convergence.

The dataset used to construct the model consists of 270 binary and ternary compounds containing sp-- or fully occupied d--metal elements, referred to as the in-domain dataset. 
An OOD dataset comprising 40 materials is also prepared. 
Unlike the in-domain dataset, the OOD dataset includes compounds with transition metals or quaternary/pentanary elements.
The degree of distributional shift between the in-domain and OOD datasets is evaluated by applying the Kolmogorov–Smirnov test to the feature distributions, as shown in Supplementary Fig. \ref{fig:OOD-dist}.
Among the 18 features, 11 exhibit statistically significant differences ($p < 0.01$, 99\% confidence level), confirming that the OOD dataset lies outside the feature distribution of the in-domain dataset.
This OOD dataset is used to test the hypothesis that an XML-guided compact model would achieve better generalization for chemically distinct systems than the more complex pristine model, which may be overfitted to the in-domain distribution.
The materials are selected from the Materials Project Database\cite{MPD} through arbitrary sampling, ensuring computational feasibility under the following constraints: fewer than 20 atoms per unit cell and no significant challenges in electronic iteration convergence.
Feature generation and prediction objectives are determined using first-principles calculations under the same computational conditions as described by Lee $et$ $al$.\cite{lee2016GWgap}
The list and properties of the OOD dataset are provided in Supplementary Table \ref{tab:OODraw}.

\subsection*{\label{subsec_method_pfi}Permutation feature importance (PFI)}
PFI\cite{pfi} is a global XML method used to quantify the contribution of a feature to the predictions of the model.
It assesses the feature importance by independently shuffling each feature and measuring the impact on the predictive performance of the model. 
The PFI score for the $j$-th feature is defined as the increase in the predictive error of the model when the feature is permuted, relative to the error of the original model.
In this study, the PFI scores are computed by shuffling features in the test dataset and evaluating the corresponding increase in the RMSE.

\subsection*{\label{subsec_method_shap}SHapley Additive exPlanation (SHAP)}

SHAP\cite{shap1} is an XML method that explains the prediction of individual samples by attributing the model's output to different features.
Rooted in cooperative game theory’s principle of fair distribution, SHAP assigns an importance value to each feature, quantifying its contribution to a specific prediction. 
Conceptually, SHAP values average a feature’s marginal contribution over all possible feature coalitions.

SHAP represents predictions via an additive explanation model:
\begin{equation}
    g(x') \;=\; \phi_{0} \;+\; \sum_{j=1}^{n_x} \phi_j \, x'_j, 
    \qquad x' \in \{0,1\}^{n_x},
\end{equation}
\noindent where $n_x$ is the number of input features in the model being explained, $x'$ is a binary coalition vector indicating feature presence ($x'_j=1$) or absence ($x'_j=0$), $\phi_j$ is the SHAP value of feature $j$, and $\phi_0$ is the base value, corresponding to the expected prediction over the background data.
For a coalition $x'$, features with $x'_j=0$ are marginalized over the background distribution, while those with $x'_j=1$ are conditioned on their observed values; aggregating over all coalitions (all $x'_j=1$) yields SHAP values that satisfy local accuracy as follows: 
\begin{equation}
    \hat f(x) \;=\; g(\mathbf{1}) \;=\; \phi_0 + \sum_{j=1}^{n_x}\phi_j .
\end{equation}

The SHAP value for the $j$-th feature is defined as
\begin{equation}
    \phi_j \;=\; \sum_{S \subseteq F \setminus \{j\}}
    \frac{|S|!\,\bigl(|F|-|S|-1\bigr)!}{|F|!}
    \Bigl[\,\hat f_{S\cup\{j\}}\!\bigl(x_{S\cup\{j\}}\bigr)
          - \hat f_{S}\!\bigl(x_{S}\bigr)\Bigr],
    \quad \text{with } |F| = n_x,
\end{equation}
\noindent where $F$ is the full set of features considered at this stage, and the sum runs over all subsets $S \subseteq F\setminus\{j\}$, excluding $j$--th feature.
Here, $\hat f_{S}(x_S)$ denotes the model prediction when only the features in $S$ are present and all other features are integrated out using the background distribution, and $\hat f_{S\cup\{j\}}(x_{S\cup\{j\}})$ is defined analogously.

SHAP can provide both local and global explanations. 
In this study, SHAP importance, a global explanation approach, is used.
It represents feature importance by averaging the absolute SHAP values across all samples.
SHAP values are computed using the model-agnostic KernelExplainer implemented in the Python SHAP library\cite{shap2} with the standardized training set as the background dataset for each split.
For each test sample, SHAP values are approximated by sampling up to 100 feature coalitions based on the Shapley kernel weighting scheme when estimating marginal contributions.
This setting is chosen to balance computational efficiency with the stability of the SHAP estimates.

\subsection*{\label{subsec_statistic}Significance test}
To assess whether one predictive model achieves a statistically significant improvement in the accuracy over another, the per-trial RMSE values are analyzed using paired $t$-tests.
In each trial, the same random seed is used for data splitting, and the resulting RMSE values of the two models are compared in pairs.
A 99\% confidence level is adopted: If the $p$-value exceeds 0.01, the two models are regarded as not significantly different, whereas a $p$ value less than 0.01 indicates that the predictive performance of one model is statistically superior.
This procedure ensures that observed performance differences are evaluated for consistency across the 20 randomized trials rather than being attributed to chance.

\section*{\label{sec_results} Results}

\subsection*{\label{subsec_correl}Feature correlations}
Before applying XML methods, the feature correlations were analyzed.
Supplementary  Fig. \ref{fig:corr} presents the correlation matrix of all features, and the detailed numerical values are presented in the publicly available code repository.
Pearson’s correlation coefficient ($r_p$) is an indicator of linear correlations based on actual values, whereas Spearman’s rank correlation coefficient ($r_s$) is used to evaluate monotonic relationships using ranked data points instead of actual values.
To mitigate redundancy, feature elimination was performed through a performance-based iterative procedure guided by pairwise correlations. Feature pairs were ranked according to the average of $r_p$ and $r_s$ and examined in descending order of correlation strength. 
For each pair, predictive models were reconstructed by excluding each feature individually, and the resulting prediction errors were compared using a paired $t$-test. If the exclusion of a feature led to a statistically significant increase in prediction error, the feature was retained and its correlated counterpart was removed. If neither exclusion caused a significant change, the feature with lower correlation to the prediction objective ($E_\mathrm{g}^\mathrm{GW}$) was removed.
This process was repeated iteratively while monitoring the prediction error relative to the pristine \textit{18-feature} model. The elimination procedure was terminated once further removal resulted in a statistically significant increase in prediction error, indicating that both features in the pair were required to maintain predictive performance.  
Consequently, seven features were eliminated from the original \textit{18-feature set} in the order: $\bar{m}$, $\sigma(Z)$, $E_\mathrm{g}^\mathrm{mBJ}$, $\bar{p}$, $\bar{I}$, $\sigma(I)$, and $V$.
The resulting feature set corresponds to a pairwise correlation level of 0.8, which arises from the performance-based elimination process.
The detailed elimination order is presented in Supplementary Table \ref{tab:feature_elimination}.

To further assess multicollinearity beyond pairwise correlation coefficients, variance inflation factors (VIF), which quantify how strongly a feature can be linearly explained by the remaining features, were additionally computed as a diagnostic measure.
The original \textit{18-feature} set exhibited extremely high VIF values, indicating severe multicollinearity.
As correlated features were sequentially removed according to the adopted criterion, the maximum VIF decreased substantially (see Supplementary Table \ref{tab:feature_elimination}), and the resulting \textit{11-feature set} showed all VIF values below 10.
This confirms that multicollinearity was effectively mitigated through the correlation-based filtering.

The removal of these features improves the consistency in the XML analysis, as further expounded in the Discussion section.

\subsection*{\label{subsec_xml_result}XML results and important feature selection}
The reduced \textit{11-feature set} was used to construct a predictive model under the same conditions used for the original \textit{18-feature set}.  
XML methods, PFI and SHAP, were then applied to assess the feature importance.

\begin{figure*}[hb]
 \begin{center}
  \includegraphics[width=0.99\linewidth]{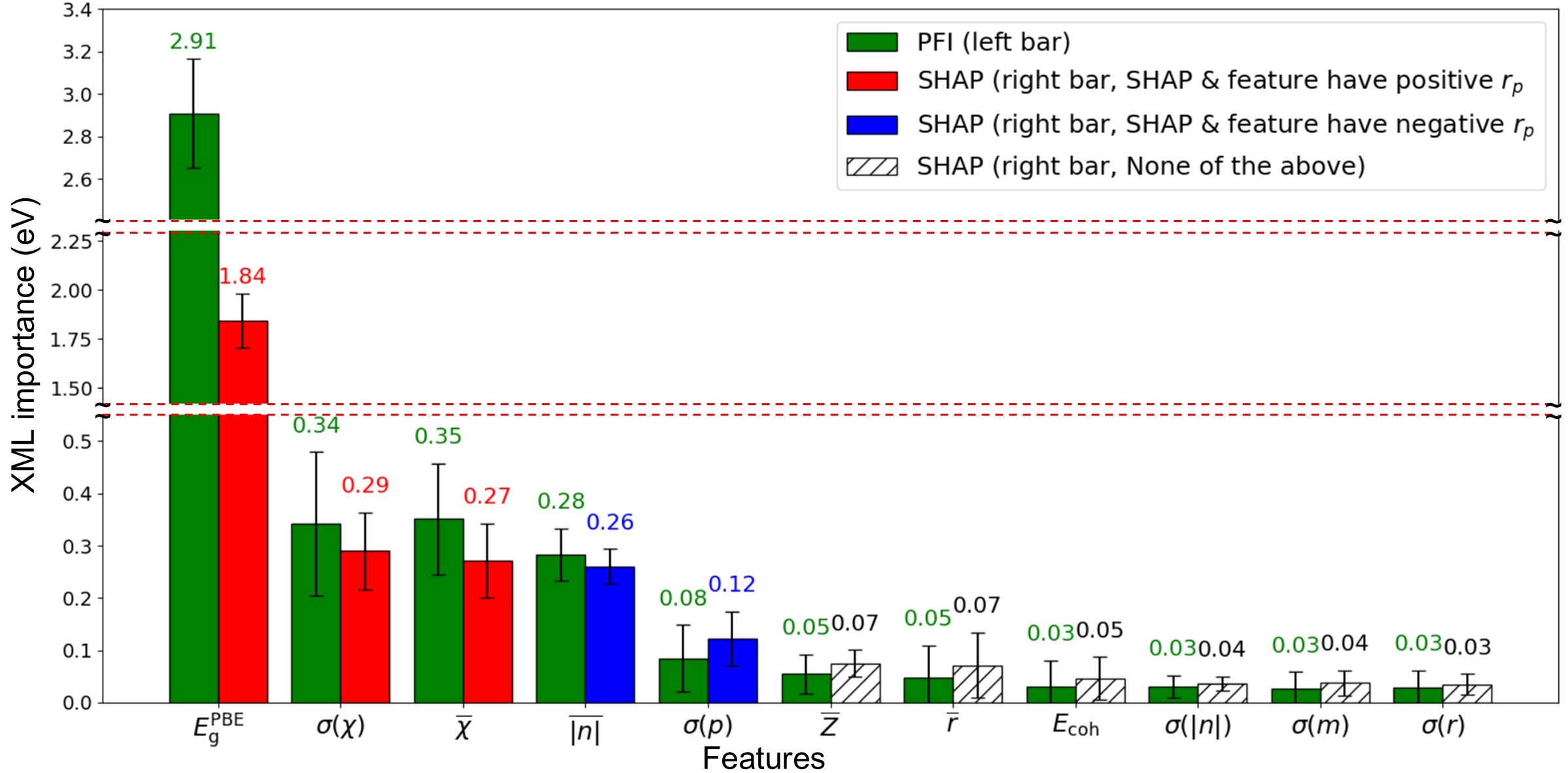}
 \caption{
 XML importance scores of SVR model for predicting $E_\textrm{g}^{\textrm{GW}}$ using \textit{11-feature set}.
 For each feature, the paired bars represent the PFI (left bar) and SHAP importance (right bar).
 The PFI score is defined as the increase in RMSE on the test dataset after randomly shuffling the values of a given feature, while keeping the model trained on the corresponding training dataset.
 The SHAP importance score is calculated as the mean of the absolute values of the individual SHAP values from the test dataset.
 The bars are colored red or blue when the sign of $r_p$ between the SHAP values and the corresponding feature values is consistently positive or negative, respectively, across the 20 predictive models.
 Hatched bars indicate that the sign varies among the models.
 The error bars represent one standard deviation of the XML importance scores across the predictive models constructed using 20 different data selections.
 Features are ordered in descending order according to their average PFI and SHAP importance scores.
 }
\label{fig:svrpfishap}
\end{center}
\end{figure*}

Figure \ref{fig:svrpfishap} (left bar) presents the PFI scores for the prediction of $E_\textrm{g}^{\textrm{GW}}$ using the SVR model.
Among the features, $E_\textrm{g}^\textrm{PBE}$ exhibited the highest PFI score (2.91 eV), indicating its dominant influence on the predictions.
The next most important features—--$\bar{\chi}$, $\sigma(\chi$), and $\mid$$\bar{n}$$\mid$--—had PFI scores ranging from 0.28 to 0.35 eV.
In contrast, the remaining features, beginning with $\sigma(p$), had relatively low PFI scores (0.03--0.08 eV), suggesting minimal impact on the performance of the model. 
A higher PFI score indicates a greater contribution to the predictive accuracy, whereas near-zero scores suggest that variations in the values of the feature do not significantly affect the predictions, rendering the feature less influential.

Figure \ref{fig:svrpfishap} (right bar) also presents the SHAP importance scores, which provide a quantitative indicator of the contribution of each feature to the prediction of $E_\textrm{g}^{\textrm{GW}}$ using the SVR model with the \textit{11-feature set}.
Among the features, $E_\textrm{g}^\textrm{PBE}$ had the highest SHAP importance score (1.84 eV), followed by $\sigma(\chi$), $\bar{\chi}$, and $\mid$$\bar{n}$$\mid$ with values ranging from 0.26 to 0.29 eV. 
The remaining features, starting from $\sigma(p$), exhibited lower SHAP importance scores (0.03--0.12 eV). 
These findings align with the results of the PFI analysis.

To further investigate the directional influence of the features on the predictions, $r_p$ was analyzed for the feature values and SHAP values using the test dataset.
A positive $r_p$ indicates that an increase in the feature contributes to an increase in the SHAP value, whereas a negative $r_p$ suggests the opposite effect.
The five most influential features ($E_\textrm{g}^\textrm{PBE}$, $\sigma(\chi$), $\bar{\chi}$, $\mid$$\bar{n}$$\mid$, and $\sigma(p$)) consistently retained the same $r_p$ sign across 20 different data selections.
Specifically, increases in $E_\textrm{g}^\textrm{PBE}$, $\sigma(\chi$), and $\bar{\chi}$ led to higher SHAP values, whereas increases in $\mid$$\bar{n}$$\mid$ and $\sigma(p$) decreased the SHAP values. 
The remaining features exhibited inconsistent $r_p$ signs and have relatively lower SHAP-importance scores.
This trend is further visualized in Supplementary Fig. \ref{fig:beeswarm}, which includes beeswarm plots illustrating the relationship between the features and SHAP values.

To assess the impact of feature selection, the feature importance scores from both XML methods were averaged and the features were ranked in descending order of importance, as shown on the horizontal axes of Fig. \ref{fig:svrpfishap}. 
Based on these rankings, ``\textit{$n_x$-feature sets}'' were constructed, where $n_x$ represents the number of selected features ($n_x$ = 2 to 11). 
Predictive models were then developed using these progressively refined feature sets, in accordance with their XML-derived importance rankings.

\subsection*{\label{subsec_performance}Dependence of reduced-feature set on predictive performance}

\begin{figure*}[!ht]
 \begin{center}
    \includegraphics[width=0.70\linewidth]{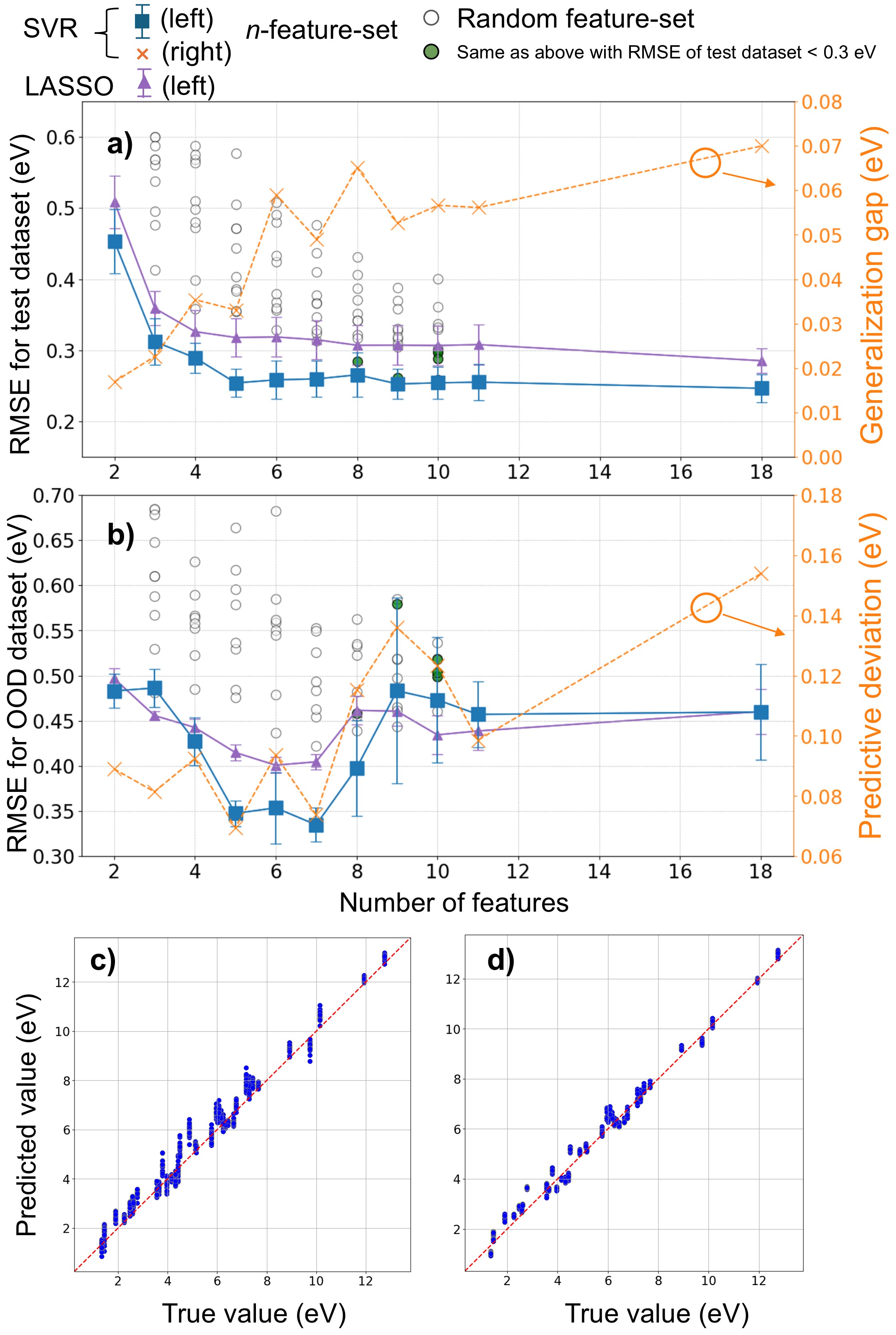}
 \caption{
  SVR models for $E_\textrm{g}^\textrm{GW}$ prediction using various feature sets.
 (\textbf{a}) Dependence of the RMSE for the test in-domain dataset (cyan rectangles) and generalization gap (orange $\times$, right vertical axis) on the number of features selected based on XML importance scores.
(\textbf{b}) Dependence of the RMSE for the OOD dataset (cyan rectangles) and the predicted value deviations (orange $\times$, right vertical axis) on the number of features selected based on the XML importance scores using 20 different data selections. 
The values at $n_x$ = 18 correspond to the pristine model.
Error bars indicate one standard deviation across the predictive models using 20 different data selections.
For $n_x$ = 3 to 10, ten predictive models with random feature sets, including $E_\textrm{g}^\textrm{PBE}$, were constructed, represented by empty circles.
In addition, models with low RMSE values for the in-domain test dataset (<0.30 eV) are represented by green circles in panels (\textbf{a}) and (\textbf{b}).
Values exceeding the range of the vertical axis are not displayed.
LASSO results are also shown for comparison; here the horizontal axis indicates the number of input features, and the feature order for $n_x$ = 2 to 11 is determined by ranking the coefficients in the \textit{11-feature} LASSO model by their absolute magnitude.
Parity plots for the 40 OOD data points: (\textbf{c}) Pristine model with \textit{18-feature set} and (\textbf{d}) predictive model with \textit{5-feature set}.
Each dot represents the predicted values from the predictive models with 20 different data selections.
The parity plots for the models with all the other feature sets are displayed in Supplementary Fig. \ref{fig:allfeatures-deviations}.
}
\label{fig:prediction}
\end{center}
\end{figure*}

\begin{figure*}[htp]
 \begin{center}
  \includegraphics[width=0.99\linewidth]{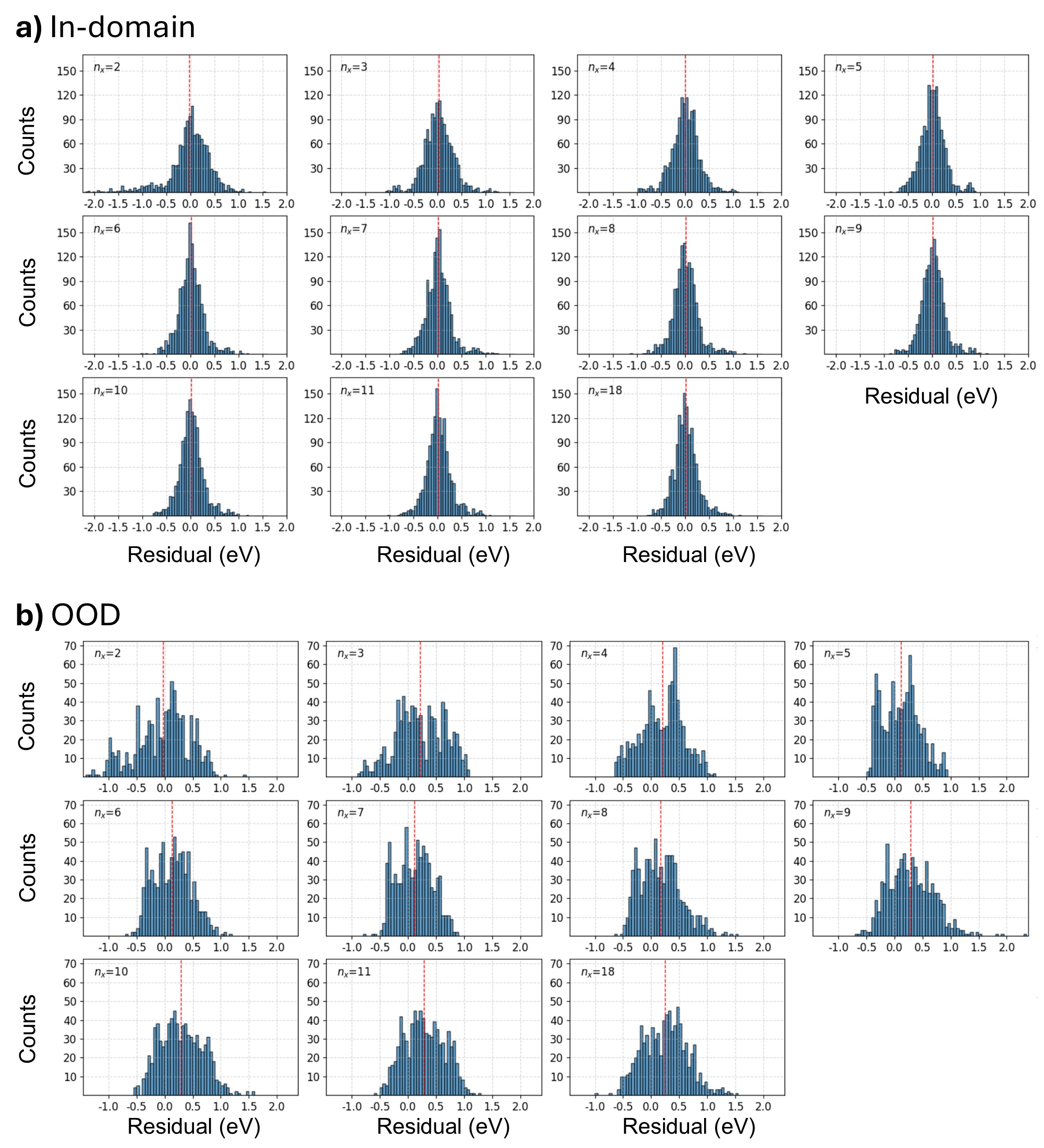}
 \caption{
 Comparison of residual distributions of SVR models for (\textbf{a}) in-domain (20 data selections $\times$ 68 test data) and (\textbf{b}) OOD (20 data selections $\times$ 40 data) data across different number of features.
  The red, dashed, vertical line indicates the mean of the residual distribution for each model.}
\label{fig:error-dist}
\end{center}
\end{figure*}

Figure \ref{fig:prediction}a shows the relationship between the prediction error (RMSE) of the test dataset and the number of selected features ($n_x$) ranked by the XML importance.
Other error-related metrics are summarized in Supplementary Table \ref{tab:svr_performance_raw} and Fig. \ref{fig:error-dist}.
A lower RMSE indicates better predictive performance.
After removing the seven highly correlated features, the predictive performance of the model employing the \textit{11-feature set} remained comparable to that of the pristine model.
As $n_x$ decreased from 11 to 5, the model maintained stable performance, with the RMSE of the test dataset ranging between 0.253 and 0.266 eV. 
However, when $n_x \leq$ 4, the predictive performance declined and the RMSE increased beyond this range, indicating that further feature reduction negatively impacts the performance of the model.

The prediction error of the training dataset gradually increased as $n_x$ decreased, as shown in Supplementary Fig. \ref{fig:trainingdata}.
A smaller $n_x$ corresponds to a simpler predictive model.
The generalization gap, defined as the difference between the training and test dataset errors, provides an indicator of overfitting and the robustness of the model, where a smaller gap suggests more consistent performance between the training and unseen data.
Notably, the generalization gap was smallest when $n_x$ = 2, although the corresponding RMSE values were higher than those obtained with models using more features.
When $n_x$ increased beyond 2, the generalization gap tended to widen, as illustrated in Fig. \ref{fig:prediction}a.

Thus far, the predictive performance of the model has been evaluated using the in-domain data.
To assess the performance in evaluating OOD data, the dependence of the prediction error on $n_x$ was also examined.
Figure \ref{fig:prediction}b shows the relationship between $n_x$ and the prediction error for the OOD dataset based on the features selected using the XML importance. 
The RMSE of the pristine model was 0.460 eV for the OOD dataset, which was significantly higher than the RMSE of 0.247 eV for the in-domain dataset.
However, predictive models using \textit{$n_x$-feature sets} ($n_x$ = 4 to 8) outperformed the pristine model in predicting the OOD data.
Notably, for $n_x$ = 5 to 7, the RMSE ranged from 0.335 to 0.354 eV, representing a reduction exceeding 0.1 eV compared to that of the pristine model.
For $n_x$ = 5, 6, and 7, the 99\% confidence intervals of the RMSE differences relative to those of the pristine models were all negative, with ranges of ($-$0.147, $-$0.075), ($-$0.145, $-$0.068), and ($-$0.157, $-$0.092), respectively.
The corresponding $p$-values were all less than $10^{-7}$, confirming that the compact models consistently and significantly outperformed the pristine model in predicting the OOD data.

To further examine the robustness of the model with respect to the size of the OOD dataset, subsampling analyses with 20, 25, 30, 35, and 40 OOD compounds were performed for the compact models with $n_x$ = 5 to 7 and for the pristine model.
The results shown in Supplementary Fig. \ref{fig:OODdatasize} confirmed that the predictive performance of the models remained stable regardless of the sample size.
Although the size of the OOD dataset is limited to 40 compounds, the statistical and subsampling analyses confirm that this number is sufficient to robustly demonstrate the superiority of the compact models.

The predictive deviations ($\bar{d}$) based on the selected data were also examined. 
The predictive deviation for the $i$-th sample ($d_i$) was calculated as the standard deviation across 20 different predictions:
\begin{equation}
d_i = \sqrt{\frac{1}{N_c} \sum_{c=1}^{N_c} (\hat{y}_{c,i} - \bar{y_i})^2},
\label{eq:deviations}
\end{equation}
where $c$ represents the attempt number for data selection, $N_c$ is the total number of attempts, $\hat{y}_{c,i}$ is the predicted value for the $i$-th sample in the $c$-th attempt, and $\bar{y_i}$ is the mean predicted value for the $i$-th sample.
The predictive deviations were then averaged across the 40 OOD data points.
The models employing the \textit{$n_x$-feature set} exhibited smaller predictive deviations than the pristine model (0.154 eV).
Notably, for $n_x$ = 2 to 7, the predictive deviations remained below 0.1 eV.

For $n_x$ = 3 to 10, random feature sets were used for comparison.
To prevent excessively high prediction errors, each random feature set included $E_\textrm{g}^\textrm{PBE}$.
In both the in-domain and OOD datasets, most random feature sets resulted in higher prediction errors than the \textit{$n_x$-feature sets} selected based on the XML importance scores.
Although some random feature sets yielded low RMSE values for the in-domain dataset (< 0.30 eV), the corresponding RMSE values for the OOD dataset exceeded 0.45 eV, as shown by the green circles in Fig. \ref{fig:prediction}b.
This suggests that these models are overfitted to the in-domain dataset and lack generalization capability for the OOD dataset.

For comparison, an interpretable linear regression baseline using LASSO was also examined.
Similar to SVR, the seven highly correlated features were first removed, and the coefficient magnitudes at $n_x$ = 11 were used as a measure of the importance for ranking the features.
Models with $n_x$ = 2 to 11 features were then constructed in descending order of importance.
As shown in Supplementary Fig. \ref{fig:lasso-11feature}, although the ranking sequence differed slightly, the top five most important features selected by LASSO were identical to those identified by SVR.
The numbers of nonzero (unshrunk) coefficients are summarized in Supplementary Fig. \ref{fig:lasso-survived}.
Notably, for $n_x$ = 2 to 8, no additional shrinkage occurred, and the input features were retained as nonzero coefficients.

Figure \ref{fig:prediction} presents a comparison of the predictive performance of the LASSO regression models with that of SVR.
Across all $n_x$, SVR consistently outperformed LASSO for treatment of the in-domain dataset.
For the OOD dataset, SVR also achieved lower errors than LASSO for $n_x$ = 4 to 8.
Similar to SVR, the predictive performance of LASSO improved as the number of features decreased compared to the pristine model, for which the feature count shrank from 18 to only about 17 on average across twenty different trials.
However, using the optimal feature count ($n_x$ = 5 to 7), the predictive performance of SVR still surpassed that of LASSO by approximately 0.05 eV, confirming the advantageous generalization performance of the former.

\section*{\label{sec_discussion} Discussion}
\subsection*{\label{subsec_discussion1}Insights for feature selection}
The proposed framework for selecting important features and reducing the dimensionality based on the XML importance offers several advantages.

First, it simplifies the predictive model, enhancing its interpretability.
This reduction in complexity also narrows the generalization gap, improving the model’s ability to predict OOD data.
Consequently, the predictive model employing only the five most important features outperforms the pristine model in predicting the OOD dataset, while the performance remains comparable to that with the in-domain dataset.
Moreover, the models with reduced-feature sets exhibit lower predictive deviations across different data selections, as shown in Fig. \ref{fig:prediction}c and Fig. \ref{fig:prediction}d.
For reference, the predictive deviations across different data selections for all the feature sets are also presented in Supplementary Fig. \ref{fig:allfeatures-deviations}.
A smaller feature set further reduces the cost of data collection and mitigates the risk of unavailable data in practical applications.

This framework provides a structured approach for determining the optimal $n_x$.
By incrementally adding features based on their XML importance ranks, the predictive performance can be monitored, eliminating the need for evaluating an overwhelming number of feature set combinations.

\begin{equation}
\textrm{Number of available combinations of feature sets} = \sum_{n_x=1}^{N_x} {C(N_x, n_x)},
\label{eq:combination}
\end{equation}

\noindent where $N_x$ is the total number of features and $C$ represents the combination function. 
For $N_x$ = 18, equation (\ref{eq:combination}) results in over 410{,}000 possible feature sets. 
In this study, 80 random feature sets were evaluated for $n_x$ = 3 to 10. 
The performance of most predictive models employing random feature sets is inferior to that of models constructed with feature sets selected based on the XML importance, for both in-domain and OOD datasets.
Some random feature sets exhibit low in-domain errors but perform poorly in predicting OOD data (green circles in Fig. \ref{fig:prediction}), underscoring that XML-guided selection does not merely identify any feature subset that appears effective in-domain, but rather ensures robust generalization for chemically distinct systems.

Given the comparable prediction error for the in-domain dataset (0.254 eV) and the reduced generalization gap compared to models with more features and the pristine model, the \textit{5-feature set} suggested by XML importance scores proves to be both sufficient and robust for predictive modeling.
Although the generalization gap is smallest for the model with \textit{2-feature set}, its RMSE value is considerably higher than those of models with more features, making them less practical for prediction.
In contrast, the model with \textit{5-feature set} maintains a low RMSE and robust generalization, thus providing the most suitable balance between accuracy and simplicity.
Moreover, this feature set achieves a sufficiently low prediction error for the OOD dataset (0.348 eV) with smaller predictive deviations compared to models with more features and the pristine model.
The \textit{5-feature} model exhibits low multicollinearity (all VIF values < 3.7), indicating that the retained features remain largely independent with minimal redundancy.

The \textit{5-feature set} consists of $E_\textrm{g}^\textrm{{PBE}}$, $\sigma(\chi)$, $\bar{\chi}$, $\mid$$\bar{n}$$\mid$, and $\sigma(p)$.
The first three features exhibit positive correlations with the prediction objective $E_\textrm{g}^\textrm{{GW}}$, whereas the latter two have negative correlations.
These correlations align with the signs observed for the features and the corresponding SHAP values. 
The $r_p$ ($r_s$) values for the features and $E_\textrm{g}^\textrm{{GW}}$, ordered on the basis of the XML importance scores, are as follows: 0.98 (0.98), 0.80 (0.81), 0.57 (0.48), $-$0.59 ($-$0.63), and $-$0.11 ($-$0.07).

Although the XML importance scores of $\bar{\chi}$ and $\mid$$\bar{n}$$\mid$ are similar to that of $\sigma(\chi)$, their correlations with the prediction objective are lower. 
Interestingly, $\sigma(p)$, which does not show a significant correlation with the prediction objective, still contributes to the improved predictive performance. 
This contribution is evident from the differences in the prediction errors for the models with the \textit{4-feature} and \textit{5-feature sets}, as shown in Fig. \ref{fig:prediction}.
The residual analysis (see Supplementary Fig. \ref{fig:residual-f5-f4}) reveals that the \textit{4-feature} model exhibits a consistent tendency toward positive residuals for compositions with larger $\sigma$($p$), indicating overestimation in these ranges. 
Including $\sigma$($p$) reduces this bias and shifts the residual distribution closer to zero. 
Notably, $\sigma$($p$) shows weak pairwise correlation with both the target $E_\textrm{g}^\textrm{{GW}}$ and the other four features ($\mid$$r_p$$\mid$ and $\mid$$r_s$$\mid$ $\leq$ 0.2), indicating that it contributes complementary rather than redundant information.

A physically grounded interpretation is that $\sigma$($p$) reflects dispersion in principal quantum numbers among constituent elements.
Since principal quantum number is associated with orbital radial extent and dielectric screening scale, period dispersion may modulate the PBE-to-GW correction across different $\sigma$($p$) ranges.
In this sense, $\sigma$($p$) does not directly correlate with the band gap itself but influences $\sigma$($p$)-dependent correction behavior that is not captured by averaged electronegativity or oxidation-number features.
This provides a physically motivated explanation for its observed contribution to improved generalization performance and illustrates how XML-based ranking can identify structurally meaningful features beyond simple linear correlations.

\begin{figure*}[htp]
 \begin{center}
  \includegraphics[width=0.99\linewidth]{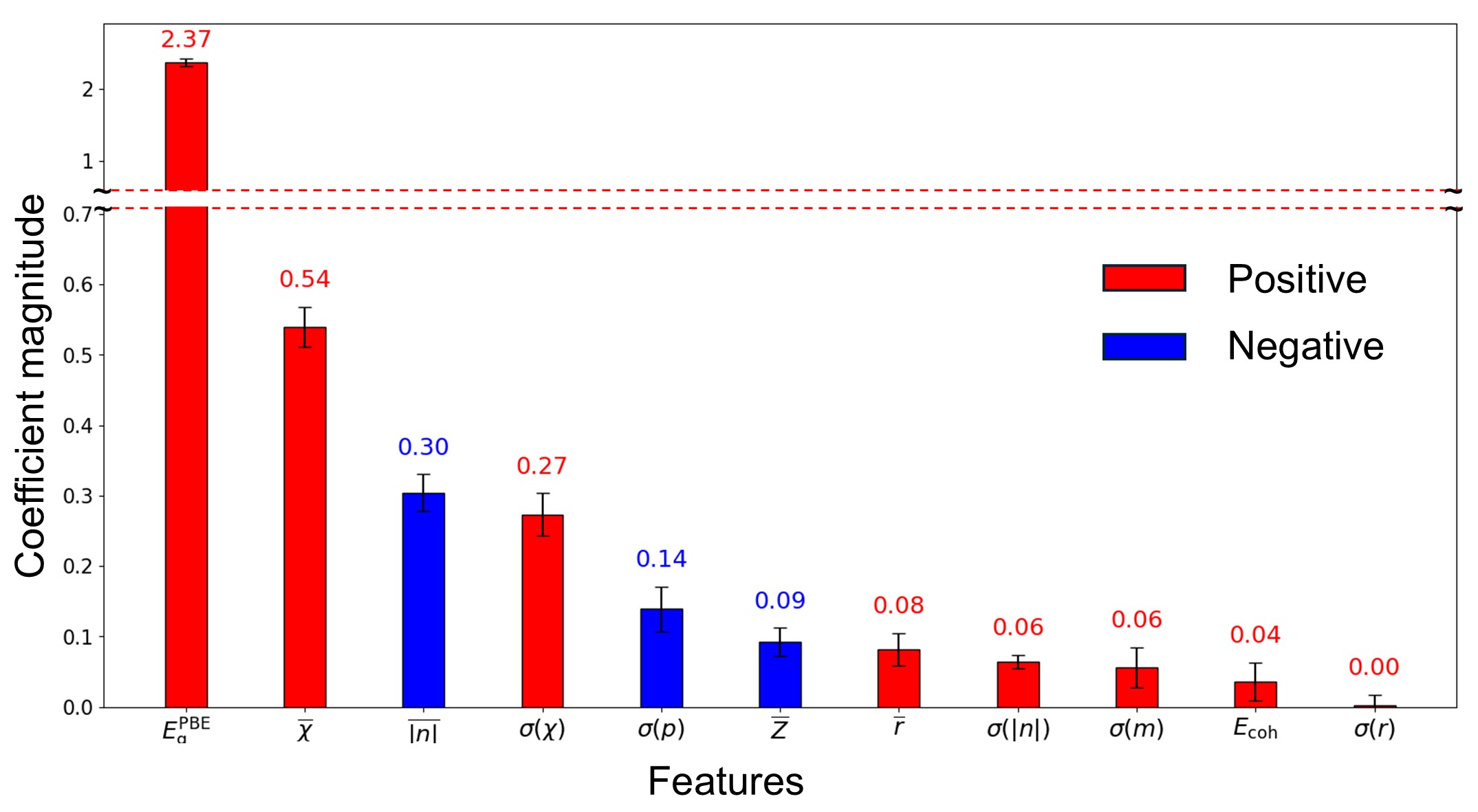}
 \caption{
  Coefficient magnitudes of LASSO regression models for predicting $E_\textrm{g}^\textrm{GW}$ using the \textit{11-feature} set. 
The absolute coefficient magnitudes serve as measures of feature importance in LASSO. 
The bars are colored red or blue when the coefficient sign remains consistently positive or negative, respectively, across the 20 predictive models.
The error bars represent one standard deviation of the coefficient magnitudes across the predictive models constructed using 20 different data selections.
The horizontal axis indicates the features ranked in descending order of their absolute coefficient magnitudes in the \textit{11-feature} model, where the same ordering is used to construct the reduced-feature models with $n_x$ = 2--11 in Fig. \ref{fig:prediction}. 
}
\label{fig:lasso-11feature}
\end{center}
\end{figure*}

The XML-based feature ranking from the SVR model was further compared with the coefficient magnitudes obtained from LASSO using the same \textit{11-feature} set (see Fig. \ref{fig:svrpfishap} and Supplementary Fig. \ref{fig:lasso-11feature}).
Although the exact ranking sequence differs slightly, the top five features identified by LASSO coincide with those obtained by the XML-guided SVR analysis.
When these five features were manually used as the input features in LASSO, the predictive performance remained comparable to that of the pristine model for the in-domain dataset and is even improved for the OOD dataset, confirming that the selected features are robust across different regression frameworks.
Moreover, although LASSO has built-in shrinkage capability, no additional shrinkage was obtained when fewer than nine input features were used (see Supplementary Fig. \ref{fig:lasso-survived}), also highlighting the usefulness and necessity of feature-selection methods based on the feature importance for different regression frameworks.

The feature set identified through XML-based importance analysis partially overlaps with that proposed by Jihad $et$ $al$.,\cite{jihad2024dft} who constructed a reduced five-feature model based on Coulombic considerations and Gaussian regression using the same compound dataset to predict $E_\textrm{g}^\textrm{{GW}}$. 
Their feature set included $E_\textrm{g}^\textrm{{PBE}}$, $\mid$$\bar{n}$$\mid$, $\bar{\chi}$, the minimum difference in electronegativity between cations and anions (min.($\Delta\chi$)), and volume$^{-1/3}$.
The latter two represent alternative formulations related to the same underlying physical quantities as $\sigma(\chi)$ and $V$ used in this study.

Although the two approaches share several common features, they are not identical.
Jihad $et$ $al$. employed physics-guided reasoning to construct a compact correction model.
In contrast, the present study employs an XML-based ranking strategy to guide feature selection within a predefined feature space.
When the feature set of Jihad $et$ $al$. is evaluated within the present validation framework, the resulting RMSE values (0.299 eV for in-domain test data and 0.369 eV for OOD) are slightly higher than those obtained using the XML-derived feature set.
The partial convergence between independently derived feature sets suggests that similar core physical information can emerge from distinct selection strategies.
The novelty of the present study lies not in proposing an entirely different feature combination, but in establishing an explicit XML-based feature selection framework that addresses correlation effects and evaluates the generalization performance of compact models.

\begin{figure*}[!ht]
 \begin{center}
  \includegraphics[width=0.95\linewidth]{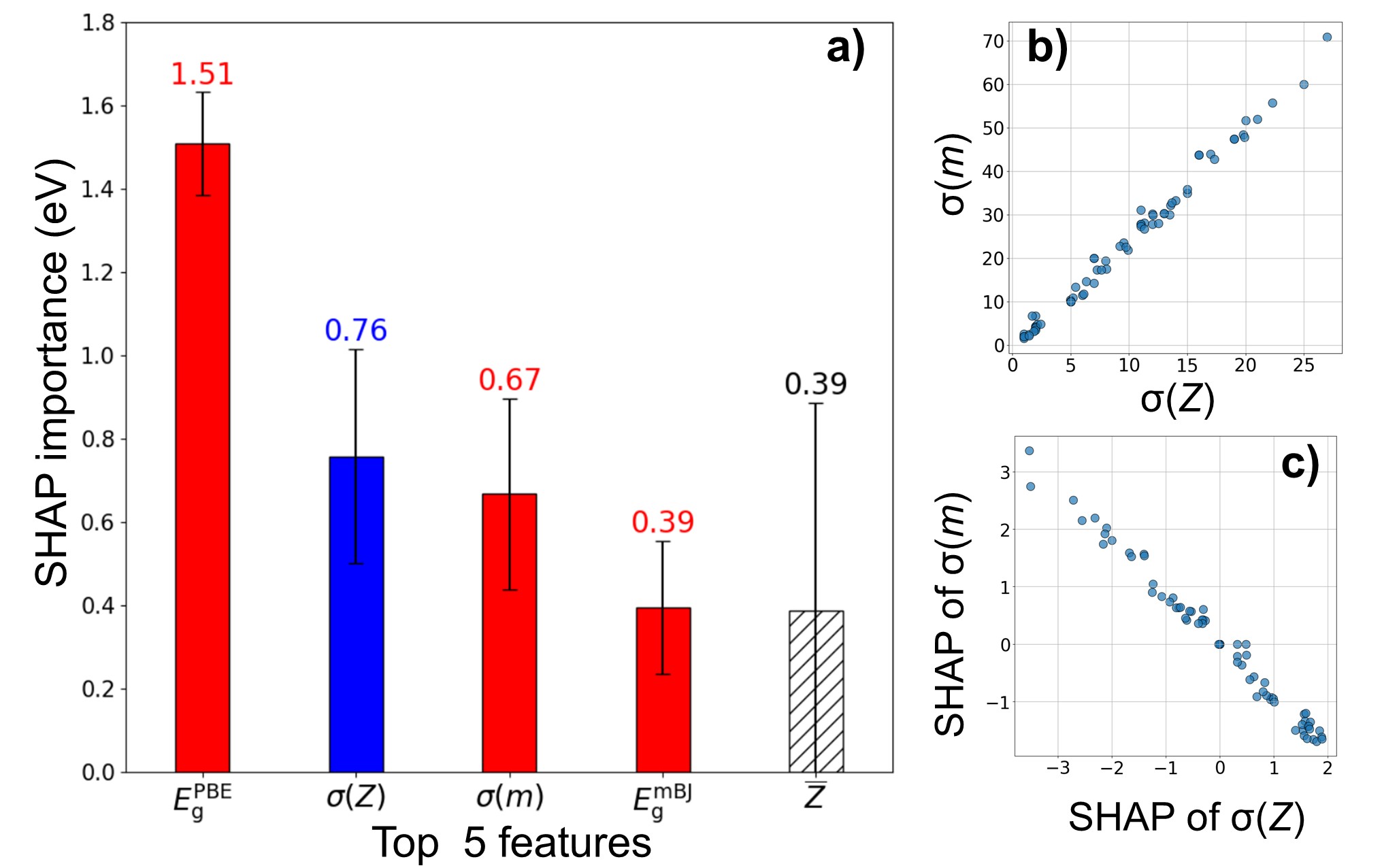}
 \caption{
  (\textbf{a}) Top five features with SHAP importance scores for SVR model for $E_\textrm{g}^\textrm{GW}$ prediction using \textit{18-feature set}.
  The SHAP importance scores for all 18 features are provided in Supplementary Fig. \ref{fig:shap17}.
  Detailed information relevant to most options, such as error bars and colors for the bar graph, is presented in Fig. \ref{fig:svrpfishap}.
  Relationships between (\textbf{b}) $\sigma(Z)$ and $\sigma(m)$, and (\textbf{c}) SHAP values for $\sigma(Z)$ and $\sigma(m)$ of test dataset. 
  }
\label{fig:wrongshap}
\end{center}
\end{figure*}
\subsection*{\label{subsec_discussion2}Correlated feature elimination prior to XML analysis}

The rationale for removing strongly correlated features before performing XML analysis is discussed here.
Figure \ref{fig:wrongshap}a presents the top five features with the highest SHAP importance scores for the pristine model using the full \textit{18-feature set}.
A complete list of the SHAP importance scores for all 18 features is provided in Supplementary Fig. \ref{fig:shap17}. 
Notably, the ranking of important features differs from that obtained with the \textit{11-feature set}.
Specifically, $\sigma(Z)$ and $\sigma(m)$ exhibit significantly high SHAP importance scores.
A similar pattern is observed in PFI analysis using the full \textit{18-feature set}, where both feature rank highly (see Supplementary Figure \ref{fig:pfi17}).
As shown in Supplementary Fig. \ref{fig:corr}, $\sigma(m)$ has a negative correlation with the prediction objective $E_\textrm{g}^\textrm{GW}$.
Nevertheless, its SHAP values are positively correlated with $\sigma(m)$ itself. 
This contradiction suggests that the presence of highly correlated features may distort the SHAP importance scores.
Figure \ref{fig:wrongshap}b illustrates the relationship between $\sigma(Z)$ and $\sigma(m)$, two features identified as important through XML analysis. 
The correlation coefficient for the relationship between these features is 0.995, indicating an extremely strong positive correlation.
Figure \ref{fig:wrongshap}c presents the distribution of the SHAP values for these two features.
Despite their strong correlation, their SHAP values exhibit opposite signs.
Such compensating behavior was not limited to the nonlinear SVR model.
A similar phenomenon was also observed in the full \textit{18-feature} LASSO model, as shown in Supplementary Fig. \ref{fig:lasso-1718}.
Even with shrinkage regularization, the number of nonzero coefficients decreased only marginally (to approximately 17).
Mirroring the behavior observed in the SHAP analysis of the SVR model, the coefficients of $\sigma(Z)$ and $\sigma(m)$ exhibited large magnitudes with opposite signs.

When the SHAP importance of $\sigma(m)$ was recalculated using the \textit{17-feature set} (excluding $\sigma(Z)$), its rank dropped from 3rd to 15th, and its importance score decreased from 0.67 to 0.03, as shown in Supplementary Fig. \ref{fig:shap17}.
A consistent trend was observed across SHAP, PFI, and LASSO analysis: when $\sigma(Z)$ was excluded from the input feature set, the PFI score of $\sigma(m)$ and its LASSO coefficient both decreased substantially (see Supplementary Figs. \ref{fig:pfi17} and \ref{fig:lasso-1718}), indicating that its apparent importance was largely driven by collinearity.
This result suggests that neither feature is inherently crucial to the predictive performance of the model; rather, their apparent importance is overestimated when both are included, as they counterbalance each other’s impact on the predicted output.
More generally, this counterbalancing effect serves as a methodological lesson: highly correlated features can artificially inflate each other’s importance scores in XML analyses, and their removal is essential for obtaining a true and reliable picture of feature contributions.
SHAP analysis assumes feature independence.\cite{xai1-book,shap1}
However, when applied in materials design, achieving complete independence is challenging because commonly used physics- and chemistry-based features often exhibit some degree of correlation. 
Therefore, strongly correlated features should be removed before performing XML analysis and regression modeling, even when using linear shrinkage methods such as LASSO, to ensure accurate and reliable results.

As discussed here, the elimination of redundant features prior to XML analysis is essential to avoid distortion in feature importance estimation. In this process, two fundamental questions arise: how far the elimination should proceed, and which feature should be removed when redundancy is identified. These questions are inherent to feature selection under multicollinearity and cannot be uniquely determined without predefined criteria.

In this study, feature elimination is guided by a performance-based stopping criterion, such that removal is continued only while predictive performance is preserved (Supplementary Table \ref{tab:feature_elimination}). As a result, the elimination process naturally converges to a feature set corresponding to a pairwise correlation level of 0.8.
This shows that the correlation level is not imposed as a predefined threshold, but emerges from a performance-constrained elimination procedure. Notably, further removal at lower correlation levels leads to a statistically significant increase in prediction error.
Therefore, a pair correlation level of 0.8 represents a practical balance between redundancy reduction and predictive stability.


In addition, when the exclusion of either feature in a highly correlated pair does not lead to a statistically significant change in prediction error, a deterministic tie-breaking rule was applied to ensure an explicit and consistent procedure. 
Specifically, between the pair, the feature with lower relevance to $E_\textrm{g}^\textrm{{GW}}$ was removed.
We verified that alternative tie-breaking strategies (e.g., removing the feature with higher VIF) led to only marginal differences in in-domain RMSE (0.252 vs. 0.255 eV) and resulted in a slightly different \textit{11-feature} configuration (e.g., removing $\overline Z$ instead of $V$; see Supplementary Table \ref{tab:feature_elimination}), while not altering the final XML-derived \textit{5-feature set}. These results indicate that the main conclusion is robust to reasonable alternative tie-breaking specifications.

From a broader methodological perspective, although the present approach and methods such as SISSO\cite{Eg_zhang, SISSO,Eg_shi} and symbolic regression\cite{symbolic1,symbolic2} can all be used in interpretable modeling contexts, they address different methodological objectives. The present study focuses on selecting a compact subset of features from a predefined and physically motivated feature space, with the goal of clarifying which originally available descriptors remain necessary after accounting for redundancy and correlation effects. By contrast, SISSO and symbolic regression construct new descriptors by combining and transforming primitive features, thereby expanding the hypothesis space beyond the original set of input variables.

This distinction is particularly relevant in the present work because the central question is not whether a newly generated descriptor can improve predictive accuracy, but whether XML-guided ranking can identify a reduced subset of the original features that preserves predictive performance while supporting improved OOD generalization. For this reason, comparison with descriptor-construction methods was not pursued in the present study, and the distinction in scope is now stated explicitly to make the intended contribution of this work clearer.

\section*{\label{sec_conclusion} Conclusion}

In this study, PFI and SHAP were combined in a systematic manner and applied to a nonlinear SVR predictive model for $E_\textrm{g}^\textrm{GW}$ to interpret and explain the predictions while identifying key features.
These XML methods provided consistent feature-importance rankings, enabling the selection of the most significant features and reducing model complexity.
By sequentially adding features based on their importance rankings, the optimal number of features was determined.
The feature importance rankings were further cross-checked against the coefficient magnitudes obtained from interpretable LASSO regression, confirming consistency between the methods.
To prevent distortion of the importance estimation, strongly correlated features were removed prior to XML analysis, and the necessity of this preprocessing step was explicitly demonstrated through a practical example.

The selected \textit{5-feature set}, based on XML importance scores, demonstrated comparable predictive performance with a smaller generalization gap for the in-domain dataset and achieved superior predictive performance for the OOD dataset compared to the more complex pristine model. 
The proposed XML-guided feature selection framework provides an explicit and practical approach for constructing compact and interpretable predictive models, which can be extended to other materials informatics applications.

\bibliography{reference}

\begin{thebibliography}{10}
\urlstyle{rm}
\expandafter\ifx\csname url\endcsname\relax
  \def\url#1{\texttt{#1}}\fi
\expandafter\ifx\csname urlprefix\endcsname\relax\def\urlprefix{URL }\fi
\expandafter\ifx\csname doiprefix\endcsname\relax\def\doiprefix{DOI: }\fi
\providecommand{\bibinfo}[2]{#2}
\providecommand{\eprint}[2][]{\url{#2}}

\bibitem{svr1}
\bibinfo{author}{Boser, B.~E.}, \bibinfo{author}{Guyon, I.~M.} \&
  \bibinfo{author}{Vapnik, V.~N.}
\newblock \bibinfo{journal}{\bibinfo{title}{A training algorithm for optimal
  margin classifiers}}.
\newblock {\emph{\JournalTitle{Proc. Fifth ann. Workshop comput. Learn.
  Theor.}}} \bibinfo{pages}{144--152} (\bibinfo{year}{1992}).

\bibitem{svr2}
\bibinfo{author}{Hearst, M.~A.}, \bibinfo{author}{Dumais, S.~T.},
  \bibinfo{author}{Osuna, E.}, \bibinfo{author}{Platt, J.} \&
  \bibinfo{author}{Sch{\"o}lkopf, B.}
\newblock \bibinfo{journal}{\bibinfo{title}{Support vector machines}}.
\newblock {\emph{\JournalTitle{IEEE Intell. Syst. Their Appl.}}}
  \textbf{\bibinfo{volume}{13}}, \bibinfo{pages}{18--28}
  (\bibinfo{year}{1998}).

\bibitem{svr3}
\bibinfo{author}{Smola, A.~J.} \& \bibinfo{author}{Sch{\"o}lkopf, B.}
\newblock \bibinfo{journal}{\bibinfo{title}{A tutorial on support vector
  regression}}.
\newblock {\emph{\JournalTitle{Stat. Comput.}}} \textbf{\bibinfo{volume}{14}},
  \bibinfo{pages}{199--222} (\bibinfo{year}{2004}).

\bibitem{deep}
\bibinfo{author}{LeCun, Y.}, \bibinfo{author}{Bengio, Y.} \&
  \bibinfo{author}{Hinton, G.}
\newblock \bibinfo{journal}{\bibinfo{title}{Deep learning}}.
\newblock {\emph{\JournalTitle{Nature}}} \textbf{\bibinfo{volume}{521}},
  \bibinfo{pages}{436--444} (\bibinfo{year}{2015}).

\bibitem{xai1-book}
\bibinfo{author}{Molnar, C.}
\newblock \emph{\bibinfo{title}{Interpretable Machine Learning}}
  (\bibinfo{year}{2022}), \bibinfo{edition}{2nd} edn.

\bibitem{xai2}
\bibinfo{author}{Zhong, X.} \emph{et~al.}
\newblock \bibinfo{journal}{\bibinfo{title}{Explainable machine learning in
  materials science}}.
\newblock {\emph{\JournalTitle{npj Comput. Mater.}}}
  \textbf{\bibinfo{volume}{8}}, \bibinfo{pages}{204} (\bibinfo{year}{2022}).

\bibitem{xai3}
\bibinfo{author}{Oviedo, F.}, \bibinfo{author}{Ferres, J.~L.},
  \bibinfo{author}{Buonassisi, T.} \& \bibinfo{author}{Butler, K.~T.}
\newblock \bibinfo{journal}{\bibinfo{title}{Interpretable and explainable
  machine learning for materials science and chemistry}}.
\newblock {\emph{\JournalTitle{Acc. Mater. Res.}}}
  \textbf{\bibinfo{volume}{3}}, \bibinfo{pages}{597--607}
  (\bibinfo{year}{2022}).

\bibitem{kohn-dft}
\bibinfo{author}{Kohn, W.} \& \bibinfo{author}{Sham, L.~J.}
\newblock \bibinfo{journal}{\bibinfo{title}{Self-consistent equations including
  exchange and correlation effects}}.
\newblock {\emph{\JournalTitle{Phys. Rev.}}} \textbf{\bibinfo{volume}{140}},
  \bibinfo{pages}{A1133--A1138} (\bibinfo{year}{1965}).

\bibitem{HSE03}
\bibinfo{author}{Heyd, J.}, \bibinfo{author}{Scuseria, G.~E.} \&
  \bibinfo{author}{Ernzerhof, M.}
\newblock \bibinfo{journal}{\bibinfo{title}{Hybrid functionals based on a
  screened {C}oulomb potential}}.
\newblock {\emph{\JournalTitle{J. Chem. Phys.}}}
  \textbf{\bibinfo{volume}{118}}, \bibinfo{pages}{8207--8215}
  (\bibinfo{year}{2003}).

\bibitem{GW}
\bibinfo{author}{Fuchs, F.}, \bibinfo{author}{Furthm{\"u}ller, J.},
  \bibinfo{author}{Bechstedt, F.}, \bibinfo{author}{Shishkin, M.} \&
  \bibinfo{author}{Kresse, G.}
\newblock \bibinfo{journal}{\bibinfo{title}{Quasiparticle band structure based
  on a generalized {K}ohn-{S}ham scheme}}.
\newblock {\emph{\JournalTitle{Phys. Rev. B}}} \textbf{\bibinfo{volume}{76}},
  \bibinfo{pages}{115109} (\bibinfo{year}{2007}).

\bibitem{lee2016GWgap}
\bibinfo{author}{Lee, J.}, \bibinfo{author}{Seko, A.},
  \bibinfo{author}{Shitara, K.}, \bibinfo{author}{Nakayama, K.} \&
  \bibinfo{author}{Tanaka, I.}
\newblock \bibinfo{journal}{\bibinfo{title}{Prediction model of band gap for
  inorganic compounds by combination of density functional theory calculations
  and machine learning techniques}}.
\newblock {\emph{\JournalTitle{Phys. Rev. B}}} \textbf{\bibinfo{volume}{93}},
  \bibinfo{pages}{115104} (\bibinfo{year}{2016}).

\bibitem{PBE}
\bibinfo{author}{Perdew, J.~P.}, \bibinfo{author}{Burke, K.} \&
  \bibinfo{author}{Ernzerhof, M.}
\newblock \bibinfo{journal}{\bibinfo{title}{Generalized gradient approximation
  made simple}}.
\newblock {\emph{\JournalTitle{Phys. Rev. Lett.}}}
  \textbf{\bibinfo{volume}{77}}, \bibinfo{pages}{3865--3868}
  (\bibinfo{year}{1996}).

\bibitem{Eg_obada}
\bibinfo{author}{Obada, D.~O.} \emph{et~al.}
\newblock \bibinfo{journal}{\bibinfo{title}{Explainable machine learning for
  predicting the band gaps of {ABX}$_3$ perovskites}}.
\newblock {\emph{\JournalTitle{Mater. Sci. Semicond. Process.}}}
  \textbf{\bibinfo{volume}{161}}, \bibinfo{pages}{107427}
  (\bibinfo{year}{2023}).

\bibitem{Eg_choubisa}
\bibinfo{author}{Choubisa, H.} \emph{et~al.}
\newblock \bibinfo{journal}{\bibinfo{title}{Interpretable discovery of
  semiconductors with machine learning}}.
\newblock {\emph{\JournalTitle{npj Comput. Mater.}}}
  \textbf{\bibinfo{volume}{9}}, \bibinfo{pages}{117} (\bibinfo{year}{2023}).

\bibitem{pfi}
\bibinfo{author}{Fisher, A.}, \bibinfo{author}{Rudin, C.} \&
  \bibinfo{author}{Dominici, F.}
\newblock \bibinfo{journal}{\bibinfo{title}{All models are wrong, but many are
  useful: Learning a variable's importance by studying an entire class of
  prediction models simultaneously.}}
\newblock {\emph{\JournalTitle{J. Mach. Learn. Res.}}}
  \textbf{\bibinfo{volume}{20}}, \bibinfo{pages}{1--81} (\bibinfo{year}{2019}).

\bibitem{Eg_zhang}
\bibinfo{author}{Zhang, L.} \emph{et~al.}
\newblock \bibinfo{journal}{\bibinfo{title}{Accurate band gap prediction based
  on an interpretable {$\Delta$}-machine learning}}.
\newblock {\emph{\JournalTitle{Mater. Today Commun.}}}
  \textbf{\bibinfo{volume}{33}}, \bibinfo{pages}{104630}
  (\bibinfo{year}{2022}).

\bibitem{SISSO}
\bibinfo{author}{Ouyang, R.}, \bibinfo{author}{Curtarolo, S.},
  \bibinfo{author}{Ahmetcik, E.}, \bibinfo{author}{Scheffler, M.} \&
  \bibinfo{author}{Ghiringhelli, L.~M.}
\newblock \bibinfo{journal}{\bibinfo{title}{{S}{I}{S}{S}{O}: A
  compressed-sensing method for identifying the best low-dimensional descriptor
  in an immensity of offered candidates}}.
\newblock {\emph{\JournalTitle{Phys. Rev. Mater.}}}
  \textbf{\bibinfo{volume}{2}}, \bibinfo{pages}{083802} (\bibinfo{year}{2018}).

\bibitem{Eg_shi}
\bibinfo{author}{Shi, Y.} \emph{et~al.}
\newblock \bibinfo{journal}{\bibinfo{title}{Interpretable machine learning for
  stability and electronic structure prediction of {J}anus {I}{I}{I}--{V}{I}
  van der {W}aals heterostructures}}.
\newblock {\emph{\JournalTitle{MGE Adv.}}} \textbf{\bibinfo{volume}{2}},
  \bibinfo{pages}{e76} (\bibinfo{year}{2024}).

\bibitem{shap1}
\bibinfo{author}{Lundberg, S.~M.} \& \bibinfo{author}{Lee, S.-I.}
\newblock \bibinfo{journal}{\bibinfo{title}{A unified approach to interpreting
  model predictions}}.
\newblock {\emph{\JournalTitle{Adv. Neural Inf. Process. Syst.}}}
  \textbf{\bibinfo{volume}{30}}, \bibinfo{pages}{4768--4777}.

\bibitem{jihad2024dft}
\bibinfo{author}{Jihad, I.}, \bibinfo{author}{Anfa, M. H.~S.},
  \bibinfo{author}{Alqahtani, S.~M.} \& \bibinfo{author}{Alharbi, F.~H.}
\newblock \bibinfo{journal}{\bibinfo{title}{{D}{F}{T}-{P}{B}{E} band gap
  correction using machine learning with a reduced set of features}}.
\newblock {\emph{\JournalTitle{Comput. Mater. Sci.}}}
  \textbf{\bibinfo{volume}{244}}, \bibinfo{pages}{113153}
  (\bibinfo{year}{2024}).

\bibitem{lasso}
\bibinfo{author}{Tibshirani, R.}
\newblock \bibinfo{journal}{\bibinfo{title}{Regression shrinkage and selection
  via the lasso}}.
\newblock {\emph{\JournalTitle{Journal of the Royal Statistical Society Series
  B: Statistical Methodology}}} \textbf{\bibinfo{volume}{58}},
  \bibinfo{pages}{267--288} (\bibinfo{year}{1996}).

\bibitem{lee2016GWgap2}
\bibinfo{title}{{G}{W}gap\_predictor\_data}.
\newblock
  \bibinfo{howpublished}{\url{http://github.com/JoohwiLEE/GWgap\_predictor\_data}}.
\newblock \bibinfo{note}{Last Accessed: Apr 06 2026}.

\bibitem{mBJ}
\bibinfo{author}{Tran, F.} \& \bibinfo{author}{Blaha, P.}
\newblock \bibinfo{journal}{\bibinfo{title}{Accurate band gaps of
  semiconductors and insulators with a semilocal exchange-correlation
  potential}}.
\newblock {\emph{\JournalTitle{Phys. Rev. Lett.}}}
  \textbf{\bibinfo{volume}{102}}, \bibinfo{pages}{226401}
  (\bibinfo{year}{2009}).

\bibitem{scikit-learn}
\bibinfo{author}{{Pedregosa, F. et al.}}
\newblock \bibinfo{journal}{\bibinfo{title}{Scikit-learn: Machine learning in
  {P}ython}}.
\newblock {\emph{\JournalTitle{J. Mach. Learn. Res.}}}
  \textbf{\bibinfo{volume}{12}}, \bibinfo{pages}{2825--2830}
  (\bibinfo{year}{2011}).

\bibitem{MPD}
\bibinfo{author}{Jain, A.} \emph{et~al.}
\newblock \bibinfo{journal}{\bibinfo{title}{Commentary: The materials project:
  A materials genome approach to accelerating materials innovation}}.
\newblock {\emph{\JournalTitle{Appl. Phys. Lett. Mater.}}}
  \textbf{\bibinfo{volume}{1}}, \bibinfo{pages}{011002} (\bibinfo{year}{2013}).

\bibitem{shap2}
\bibinfo{title}{{S}{H}{A}{P}}.
\newblock \bibinfo{howpublished}{\url{http://github.com/shap/shap}}.
\newblock \bibinfo{note}{Last Accessed: Nov 28 2023}.

\bibitem{symbolic1}
\bibinfo{author}{Schmidt, M.} \& \bibinfo{author}{Lipson, H.}
\newblock \bibinfo{journal}{\bibinfo{title}{Distilling free-form natural laws
  from experimental data}}.
\newblock {\emph{\JournalTitle{Science}}} \textbf{\bibinfo{volume}{324}},
  \bibinfo{pages}{81--85} (\bibinfo{year}{2009}).

\bibitem{symbolic2}
\bibinfo{author}{Udrescu, S.-M.} \& \bibinfo{author}{Tegmark, M.}
\newblock \bibinfo{journal}{\bibinfo{title}{{A}{I} {F}eynman: A
  physics-inspired method for symbolic regression}}.
\newblock {\emph{\JournalTitle{Sci. Adv.}}} \textbf{\bibinfo{volume}{6}},
  \bibinfo{pages}{eaay2631} (\bibinfo{year}{2020}).

\end{thebibliography}

\section*{Data Availability}
The in-domain data can be obtained at \url{https://github.com/JoohwiLEE/GWgap\_predictor\_data}.
Other raw\text{/}processed data can be found at \url{https://github.com/ToyotaCRDL/XML\_bandgap}.

\section*{Code Availability}
The code used in this study is downloadable at \url{https://github.com/ToyotaCRDL/XML\_bandgap}.

\section*{Acknowledgements}
The authors would like to thank Enago (www.enago.jp) for the English language review.

\section*{Author Contributions}

J. L. primarily performed the simulations and prepared the first manuscript.
K. M. designed the project.
All authors discussed the results and wrote the manuscript.

\section*{Funding}
This research did not receive funding.

\section*{Additional information}

\textbf{Competing interests}
The authors declare no competing interests.
\newpage

\clearpage
\onecolumn
\beginsupplement


\begin{figure*}[h!]
 \begin{center}
  \includegraphics[width=0.99\linewidth]{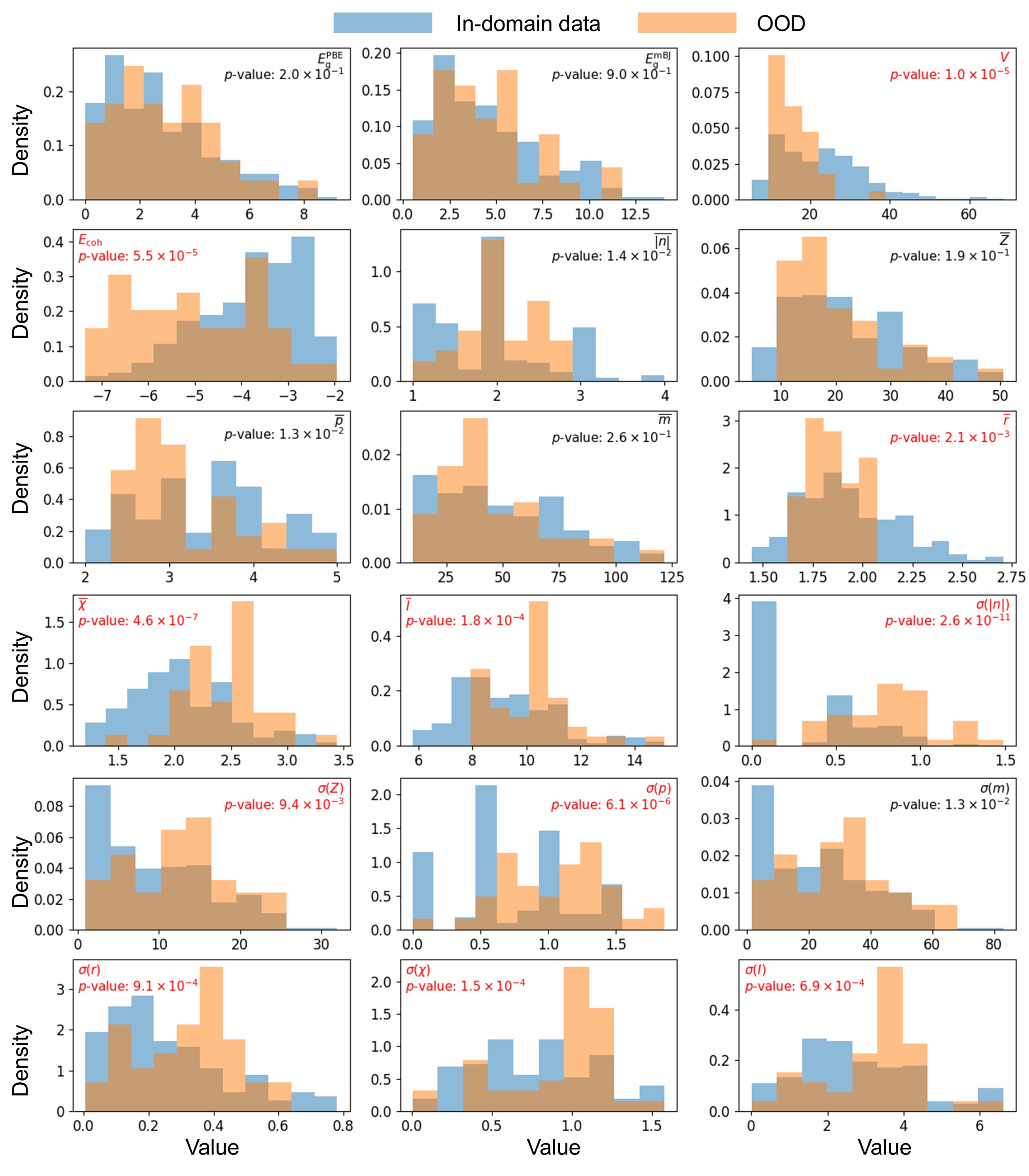}
 \caption{
 Comparison of in-domain (270 compounds composed of binary and ternary systems) and out-of-domain (OOD) datasets (40 compounds containing transition metals and/or quaternary/pentanary systems) for 18 material features.
 Each subplot shows the probability density distributions of the corresponding feature for both datasets.
The Kolmogorov–Smirnov test is used to determine whether the two distributions originate from the same population.
 The $p$-value is shown in each panel; values smaller than 0.01, shown in red, indicate that the in-domain and OOD distributions are significantly different at the 99\% confidence level.}
\label{fig:OOD-dist}
\end{center}
\end{figure*}
\restoregeometry

\begin{figure*}[htp]
 \begin{center}
  \includegraphics[width=0.99\linewidth]{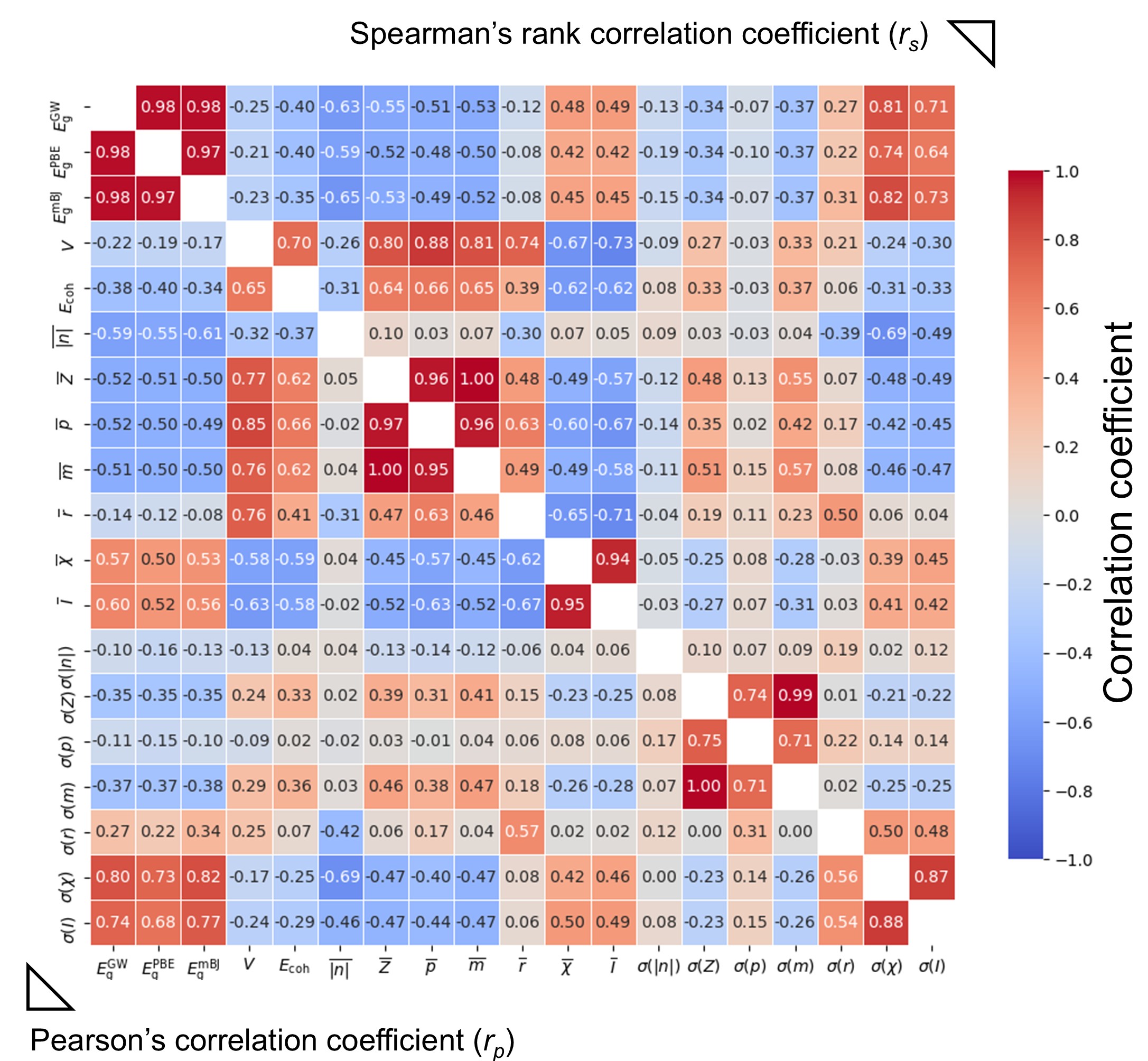}
 \caption{
  Correlation coefficients for relationship between $E_\textrm{g}^{\textrm{GW}}$ and 18 features for 270 binary and ternary inorganic compounds.
The left-lower and right-upper triangles represent $r_p$ and $r_s$, respectively.
Detailed numerical values are provided in Supplementary Data.
}
\label{fig:corr}
\end{center}
\end{figure*}

\begin{figure*}[htp]
 \begin{center}
  \includegraphics[width=0.99\linewidth]{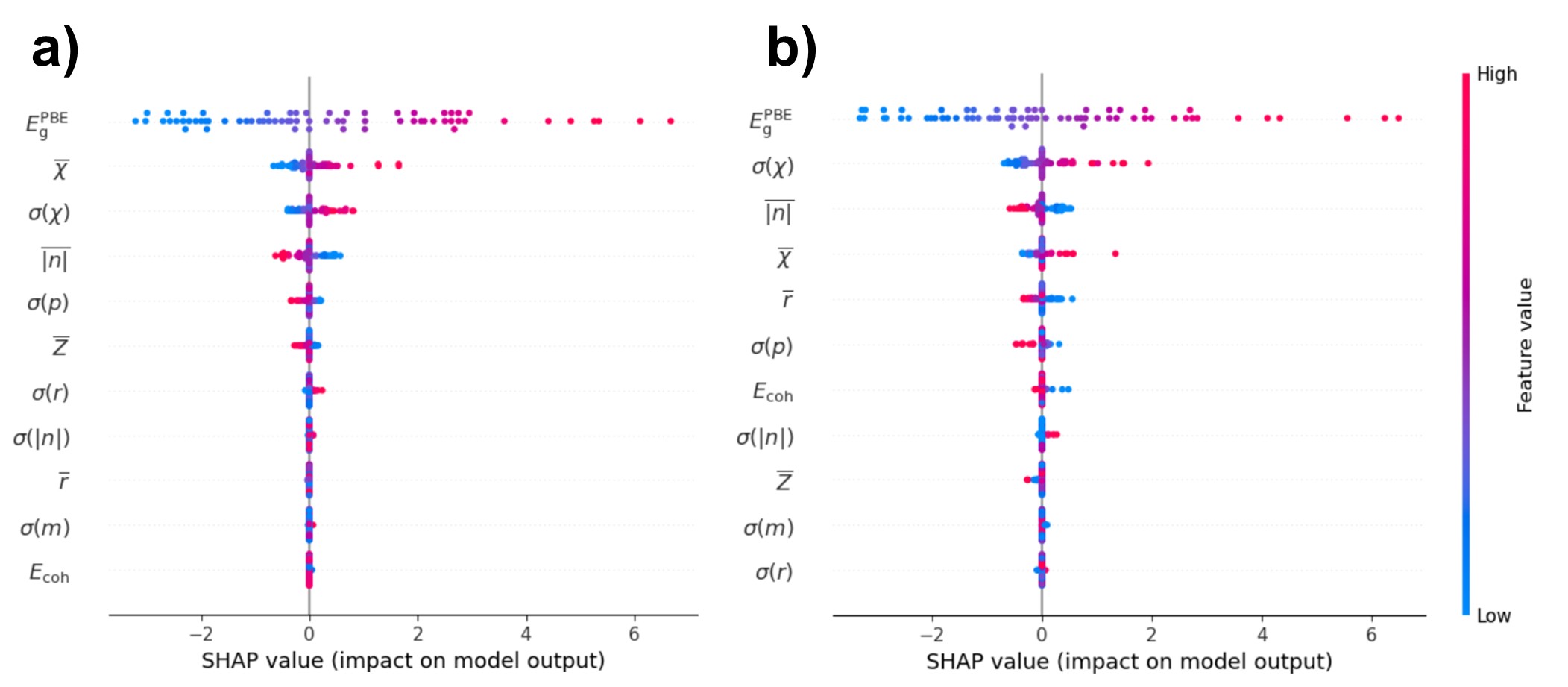}
 \caption{
 Beeswarm plots from SHAP analysis using SVR model with \textit{11-feature set} for predicting $E_\textrm{g}^{\textrm{GW}}$.
 The plots illustrate the relationships between the SHAP values and features in the test dataset, along with their distributions in panels (\textbf{a}) and (\textbf{b}), representing two different data samplings.
 Notably, for some features with low SHAP importance (indicated by small feature-distribution ranges), the impact of increases or decreases in the feature values on the SHAP values follows different trends for different samples.}
\label{fig:beeswarm}
\end{center}
\end{figure*}
\begin{figure*}[htp]
 \begin{center}
  \includegraphics[width=0.99\linewidth]{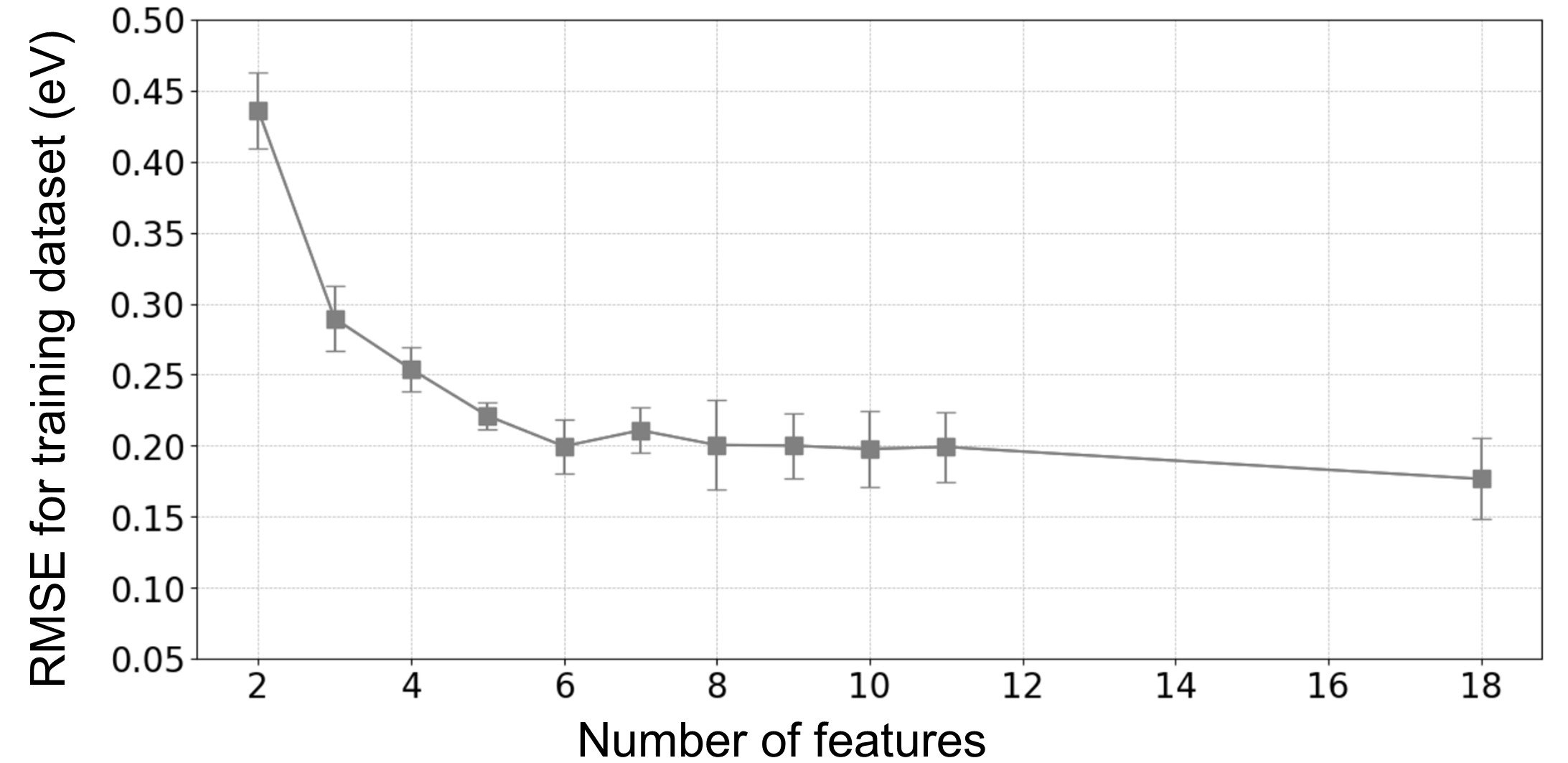}
 \caption{
  Dependence of RMSE for the training dataset of SVR model on the number of features selected based on XML importance scores.
 The value at $n_x$ = 18 corresponds to the pristine model.
 The error bars indicate one standard deviation across the predictive models with 20 different data selections.}
\label{fig:trainingdata}
\end{center}
\end{figure*}

\begin{figure*}[htp]
 \begin{center}
  \includegraphics[width=0.99\linewidth]{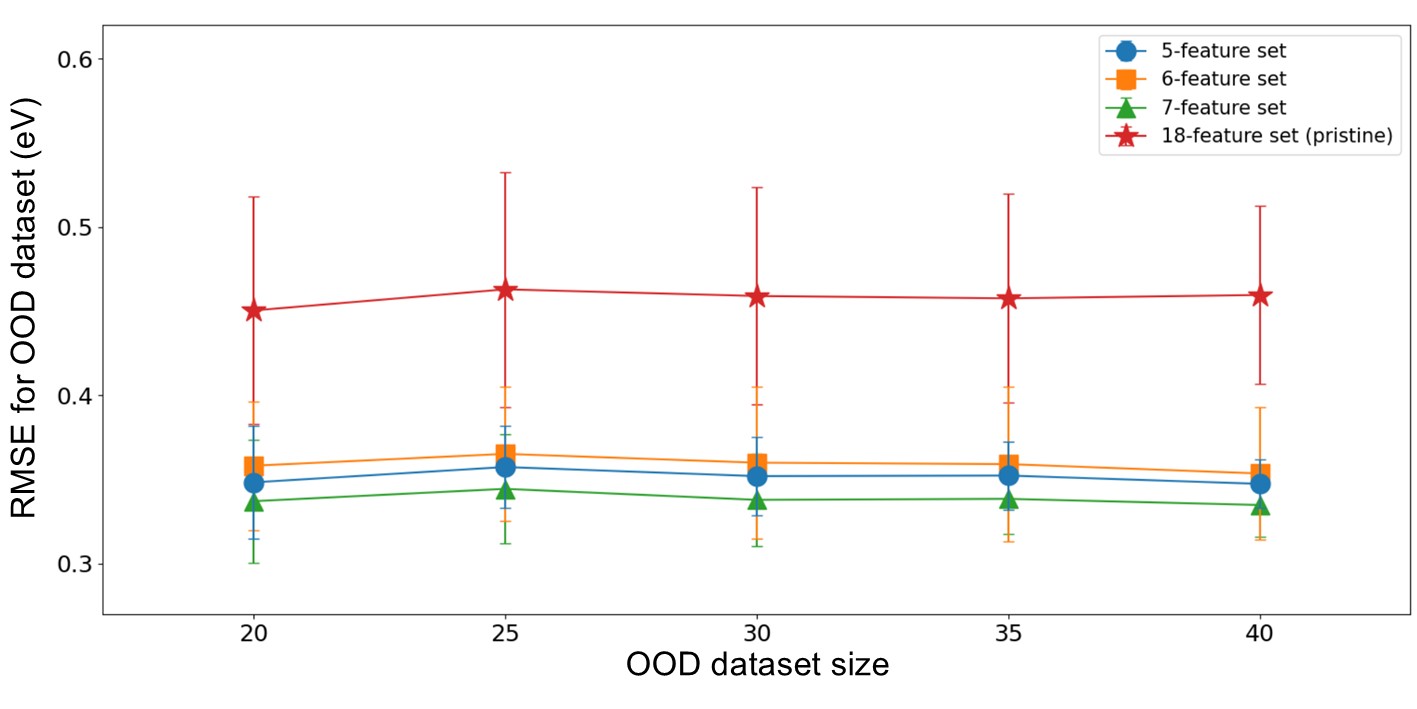}
 \caption{
  Dependence of RMSE for the OOD dataset on size of OOD dataset.
 The value at $n_x$ = 18 corresponds to the pristine model.
 The error bars indicate one standard deviation across the predictive models with 20 different data selections.}
\label{fig:OODdatasize}
\end{center}
\end{figure*}
\begin{figure*}[htp]
 \begin{center}
  \includegraphics[width=0.8\linewidth]{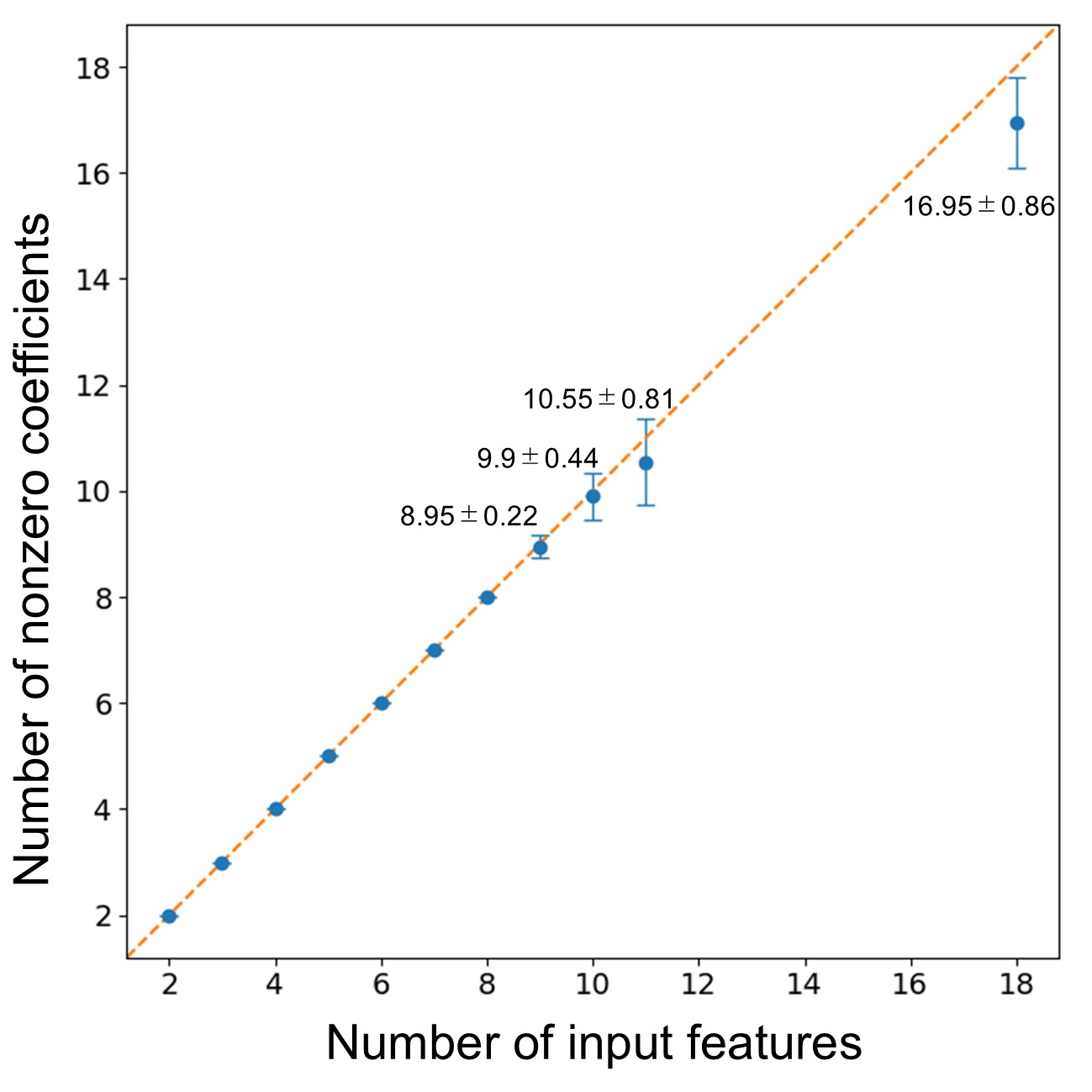}
 \caption{
  Relationship between the number of input features and the number of nonzero (unshrunk) coefficients in LASSO regression models. 
Points and error bars represent the mean and one standard deviation obtained using 20 different data selections. 
For LASSO models with $n_x \leq 11$, the input feature subsets are determined based on the feature importance ranking derived from the absolute coefficients of the \textit{11-feature} LASSO model. 
The diagonal line indicates the case where the numbers of input features and nonzero coefficients are equal.
}
\label{fig:lasso-survived}
\end{center}
\end{figure*}
\begin{figure*}[htp]
 \begin{center}
  \includegraphics[width=0.99\linewidth]{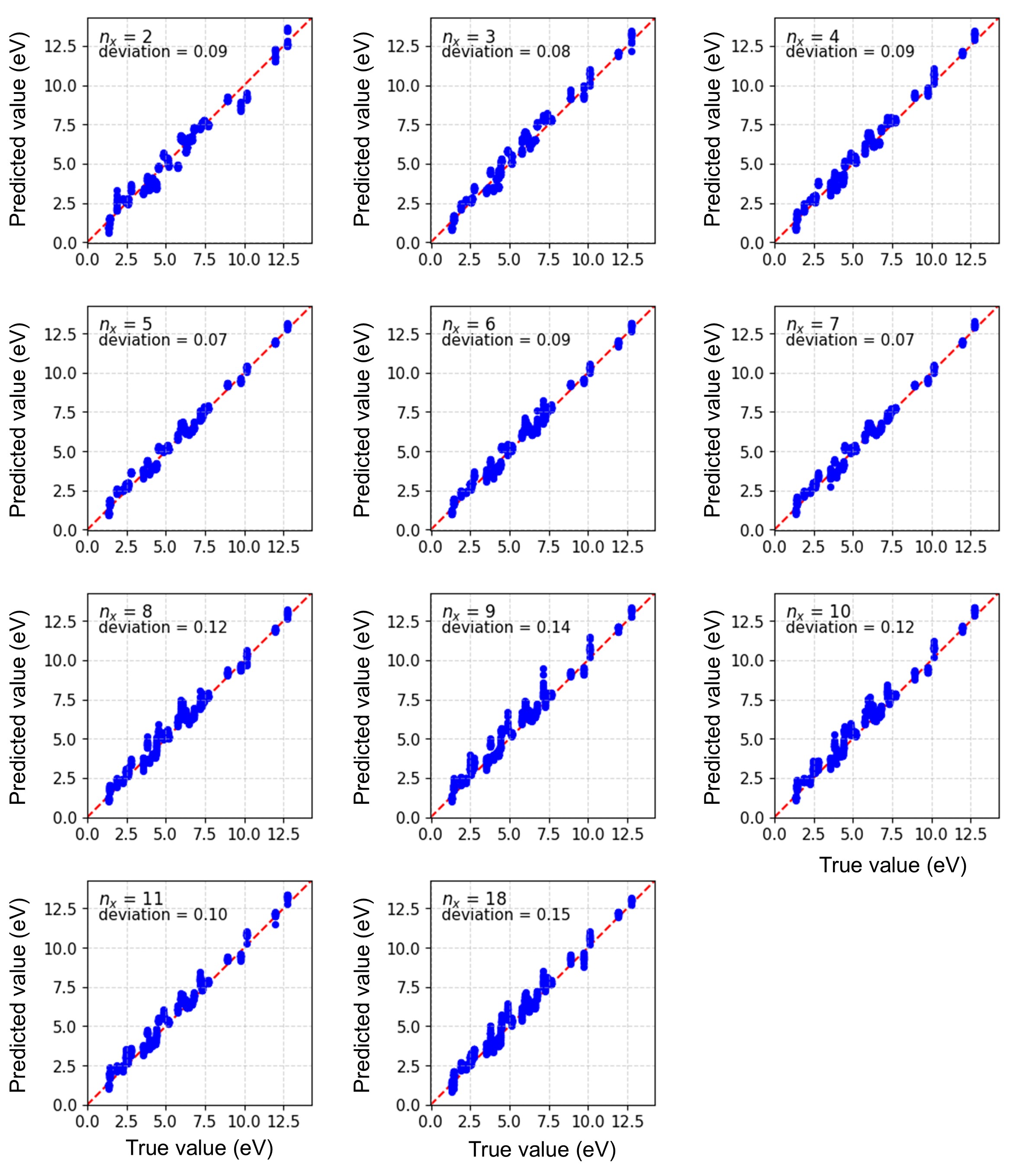}
 \caption{
  Parity plots for the 40 OOD data points using SVR models with various numbers of features. 
Each dot represents the predicted values from the predictive models with 20 different data selections.
The deviations of these predictions, obtained using equation \ref{eq:deviations}, are also shown in each subplot.
}
\label{fig:allfeatures-deviations}
\end{center}
\end{figure*}
\begin{figure*}[htp]
 \begin{center}
  \includegraphics[width=0.99\linewidth]{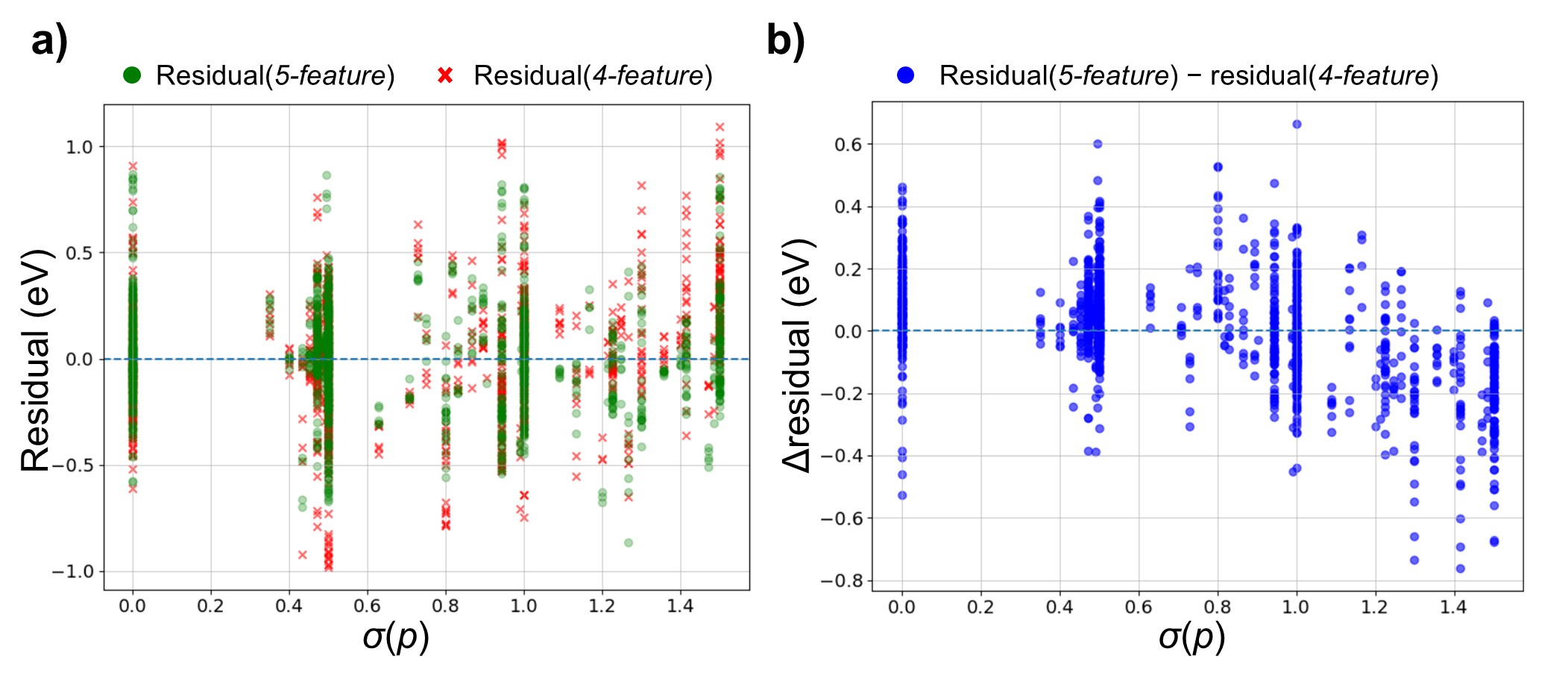}
 \caption{
  (\textbf{a}) Distribution of prediction residuals for the in-domain test datasets (68 samples $\times$ 20 trials = 1360 points) for the \textit{4-feature} (red) and  \textit{5-feature} (green) SVR models as a function of $\sigma$($p$).
  The discrete values of $\sigma$($p$) reflect compositional regimes with different degrees of period mixing.
  At larger $\sigma$($p$), the \textit{4-feature} model shows a consistent tendency toward positive residuals (overestimation), whereas inclusion of $\sigma$($p$) reduces this bias and shifts the residual distribution closer to zero.
  (\textbf{b}) Residual difference between the \textit{5-feature} and \textit{4-feature} models as a function of $\sigma$($p$).
  Negative values indicate reduction of the overestimation bias relative to the \textit{4-feature} model. 
  The bias reduction becomes more pronounced at larger $\sigma$($p$), indicating that $\sigma$($p$) modulates prediction behavior across compositional regimes rather than acting through direct linear correlation with the prediction objective.}
\label{fig:residual-f5-f4}
\end{center}
\end{figure*}

\begin{figure*}[htp]
 \begin{center}
  \includegraphics[width=0.99\linewidth]{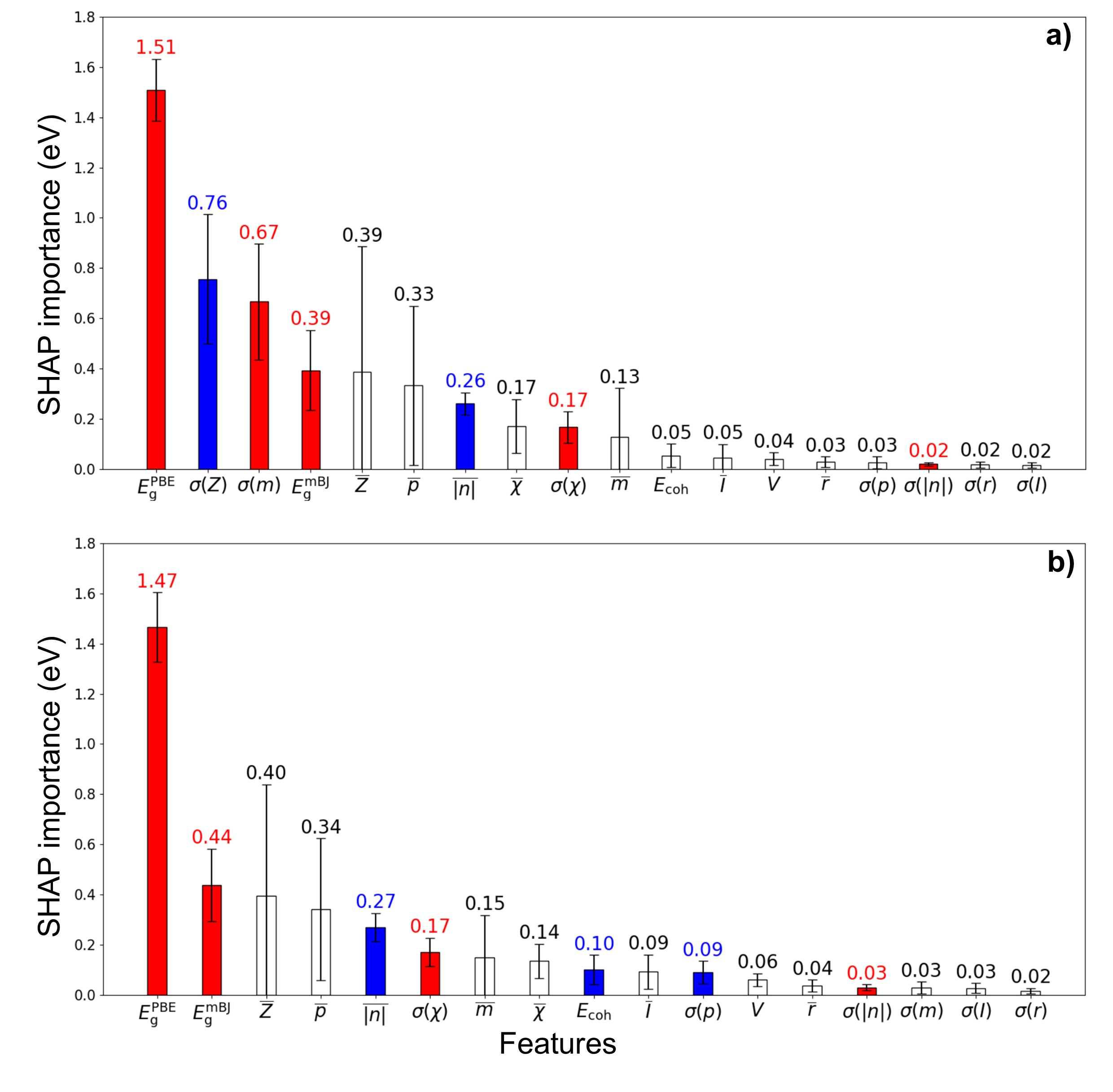}
 \caption{SHAP importance scores for SVR models for predicting $E_\textrm{g}^\textrm{GW}$ (\textbf{a}) with \textit{18-feature set} and (\textbf{b}) with \textit{17-feature set}, excluding $\sigma(Z)$.
 Detailed information relevant to most options such as error bars and colors for the bar graph are presented in Fig. \ref{fig:svrpfishap}.
}
\label{fig:shap17}
\end{center}
\end{figure*}
\begin{figure*}[htp]
 \begin{center}
  \includegraphics[width=0.99\linewidth]{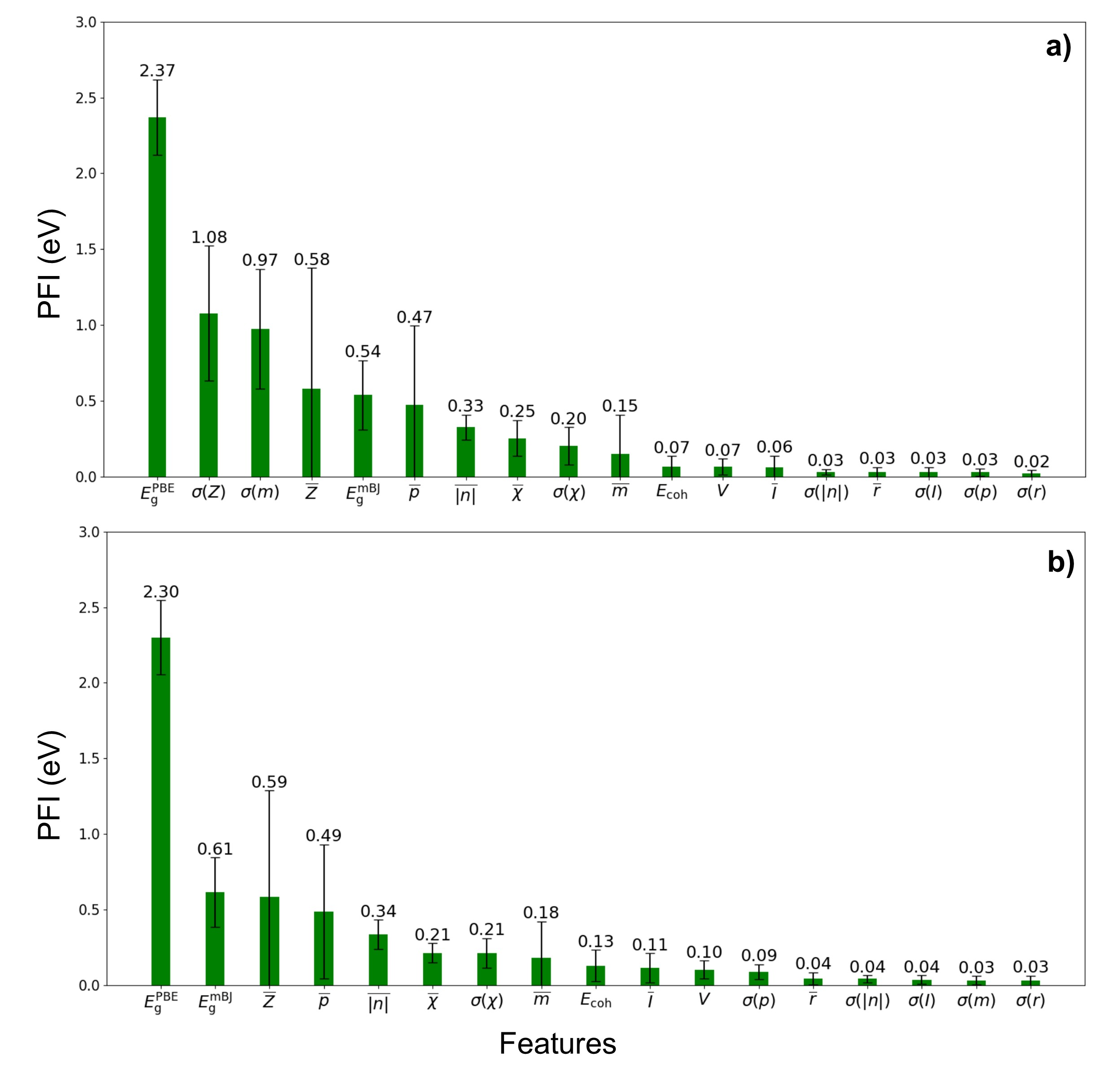}
 \caption{PFI scores for SVR models for predicting $E_\textrm{g}^\textrm{GW}$ (\textbf{a}) with \textit{18-feature set} and (\textbf{b}) with \textit{17-feature set}, excluding $\sigma(Z)$.
 Detailed information relevant to most options such as error bars for the bar graph are presented in Fig. \ref{fig:svrpfishap}.
}
\label{fig:pfi17}
\end{center}
\end{figure*}
\begin{figure*}[htp]
  \includegraphics[width=0.99\linewidth]{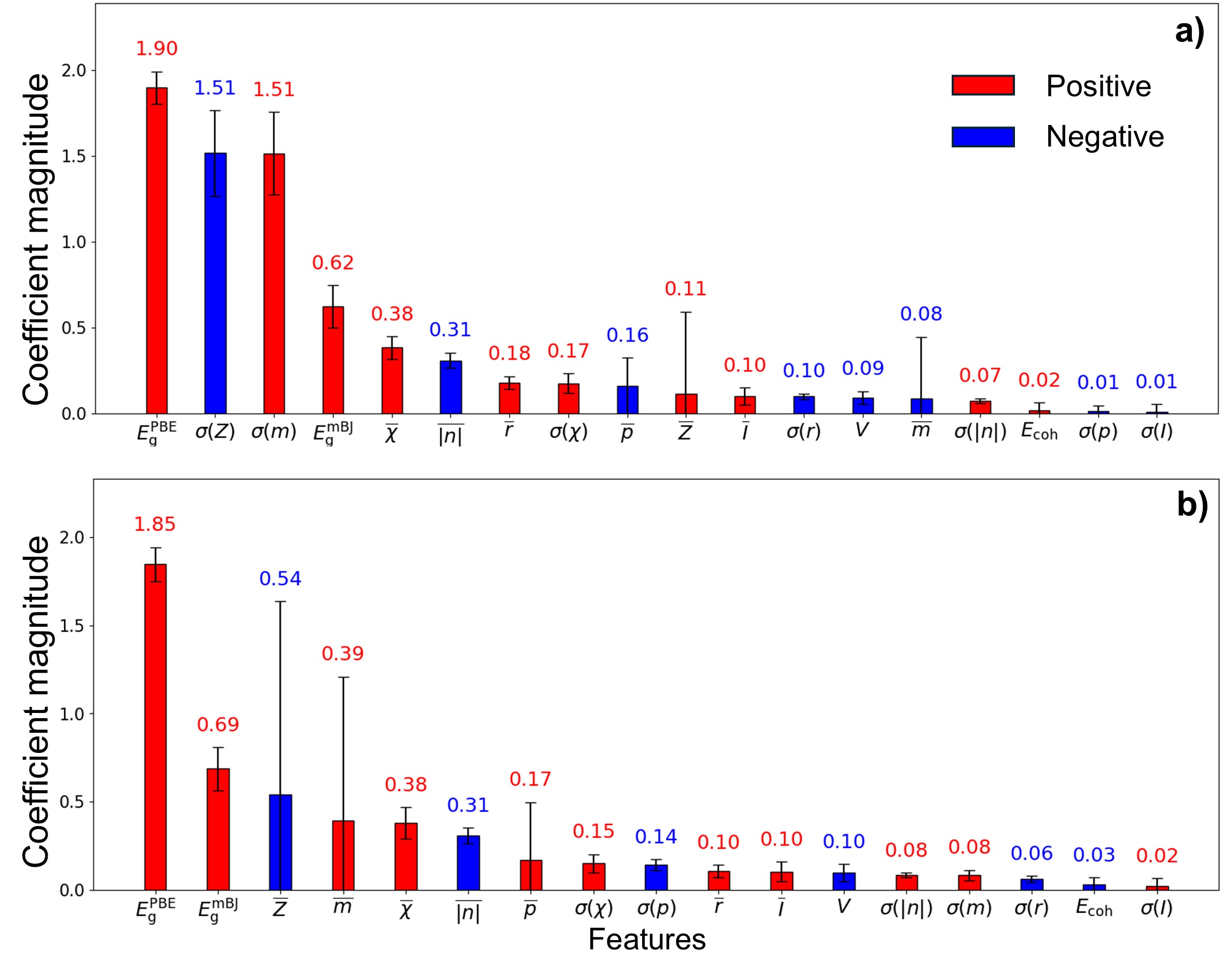}
 \caption{Coefficient magnitudes for LASSO regression models for predicting $E_\textrm{g}^\textrm{GW}$ (\textbf{a}) with \textit{18-feature set} and (\textbf{b}) with \textit{17-feature set}, excluding $\sigma(Z)$.
 Detailed information relevant to most options such as error bars and colors for the bar graph are presented in Supplementary Fig. \ref{fig:lasso-11feature}.
}
\label{fig:lasso-1718}
\end{figure*}

\clearpage
\newgeometry{left=1cm,right=1cm,top=1cm,bottom=1.5cm}

\begin{table}[tp]
  \begin{center}
  \begin{minipage}{\textwidth}
  \caption{
  Prediction objective ($E_\textrm{g}^{\textrm{GW}}$) and \textit{18-feature set} for 40 inorganic compounds in the OOD dataset.
  }\label{tab:OODraw}
  \begin{tabular*}{0.95\linewidth}{lcccccccccc}
\toprule
Material & MPD ID$^($\footnotemark[1]$^)$ & $E_\mathrm{g}^\mathrm{GW,}$$^($\footnotemark[2]$^)$ & $E_\mathrm{g}^\mathrm{PBE,}$$^($\footnotemark[2]$^)$ & $E_\mathrm{g}^\mathrm{mBJ,}$$^($\footnotemark[2]$^)$ & $V$$^($\footnotemark[3]$^)$ & $E_\mathrm{coh}$$^($\footnotemark[4]$^)$ & $\mid$$\bar{n}$$\mid$ & $\bar{Z}$ & $\bar{p}$ & $\bar{m}$\\
\midrule%
Ba$_2$MgB$_2$O$_6$ & 9259 & 7.2957 & 4.5121 & 8.1402 & 12.9682 & $-$6.2480 & 2.1818 & 16.5455 & 2.8182 & 37.8710 \\
K$_2$LiAlF$_6$ & 15549 & 11.9348 & 7.0616 & 11.1670 & 13.2148 & $-$4.5830 & 1.2000 & 10.8000 & 2.5000 & 22.6175 \\
KAlSiO$_4$ & 9480 & 7.6628 & 4.4997 & 7.2609 & 14.9860 & $-$6.0202 & 2.2857 & 11.1429 & 2.5714 & 22.5949 \\
KLi$_6$BiO$_6$ & 23582 & 3.9671 & 1.4444 & 2.4773 & 11.2193 & $-$4.0971 & 1.7143 & 12.0000 & 2.4286 & 27.5813 \\
CaTiO$_3$ & 4019 & 4.4942 & 2.4086 & 3.2974 & 11.0575 & $-$6.6312 & 2.4000 & 13.2000 & 2.8000 & 27.1916 \\
Ag$_2$GePbS$_4$ & 861942 & 2.5731 & 1.3148 & 1.9023 & 21.6331 & $-$3.3590 & 2.0000 & 34.0000 & 4.0000 & 77.9763 \\
NaAlS$_2$O$_8$ & 1210191 & 9.7422 & 5.7471 & 8.3629 & 11.5838 & $-$5.0382 & 2.0000 & 10.0000 & 2.3333 & 20.1753 \\
AgAlS$_2$ & 5782 & 3.6427 & 1.9515 & 3.0520 & 20.6793 & $-$3.8678 & 2.0000 & 23.0000 & 3.5000 & 49.7453 \\
NbCu$_3$Se$_4$ & 4043 & 2.4735 & 1.3589 & 1.5010 & 21.9054 & $-$4.0599 & 2.0000 & 33.0000 & 4.1250 & 74.9235 \\
Ba$_2$SrWO$_6$ & 18764 & 4.8743 & 2.8955 & 5.9543 & 15.7313 & $-$6.1316 & 2.4000 & 27.2000 & 3.5000 & 64.2124 \\
RbSrCO$_3$F & 863745 & 7.4352 & 4.2828 & 7.7007 & 16.4389 & $-$5.5448 & 2.0000 & 16.2857 & 2.8571 & 36.0139 \\
CaYAlO$_4$ & 1227044 & 6.1561 & 3.6202 & 5.1058 & 11.0947 & $-$6.5271 & 2.2857 & 14.8571 & 2.8571 & 31.4236 \\
Sr$_2$ScGaO$_5$ & 1105158 & 5.1163 & 2.7354 & 4.7175 & 14.1686 & $-$5.8407 & 2.2222 & 18.6667 & 3.1111 & 41.1021 \\
Sr$_2$GaSbO$_6$ & 6304 & 4.3042 & 1.5673 & 3.0119 & 12.2842 & $-$5.1016 & 2.4000 & 20.6000 & 3.1000 & 46.2713 \\
CdCu$_2$GeS$_4$ & 13982 & 1.4490 & 0.4779 & 1.0420 & 19.4042 & $-$3.4389 & 2.0000 & 25.2500 & 3.6250 & 55.0471 \\
Sr$_2$NbInO$_6$ & 20222 & 6.0101 & 3.6961 & 4.5151 & 13.4476 & $-$5.9338 & 2.4000 & 21.4000 & 3.2000 & 47.8970 \\
RbNaTiO$_3$ & 556185 & 6.6509 & 3.5750 & 4.6381 & 14.6161 & $-$5.3528 & 2.0000 & 15.6667 & 3.0000 & 34.0563 \\
Li$_2$CdGeO$_4$ & 7688 & 5.1504 & 2.5202 & 4.6497 & 11.7934 & $-$4.4159 & 2.0000 & 14.7500 & 2.6250 & 32.8801 \\
CeSiO$_4$ & 10523 & 4.4220 & 1.6060 & 1.6553 & 12.6273 & $-$6.9179 & 2.6667 & 17.3333 & 2.8333 & 38.7000 \\
CdAg$_2$I$_4$ & 1025377 & 3.5544 & 1.7747 & 2.9001 & 36.1526 & $-$1.9795 & 1.1429 & 50.5714 & 5.0000 & 119.3947 \\
Rb$_2$Li$_2$GeO$_4$ & 8450 & 6.4371 & 3.5351 & 5.8293 & 15.0273 & $-$4.2558 & 1.7778 & 16.0000 & 2.8889 & 35.7296 \\
SrTiO$_3$ & 5229 & 3.7909 & 1.8501 & 2.7358 & 11.8274 & $-$6.5718 & 2.4000 & 16.8000 & 3.0000 & 36.7000 \\
ZnCu$_2$SiTe$_4$ & 1078498 & 1.3345 & 0.0622 & 0.6795 & 26.0467 & $-$2.9551 & 2.0000 & 38.7500 & 4.3750 & 91.3709 \\
SrAgTeF & 1080438 & 2.7707 & 1.4149 & 2.5868 & 21.8604 & $-$3.8344 & 1.5000 & 36.5000 & 4.2500 & 85.5215 \\
Li$_2$ZnSnS$_4$ & 555186 & 4.1526 & 2.2812 & 3.2152 & 21.1218 & $-$3.4531 & 2.0000 & 18.7500 & 3.1250 & 40.7980 \\
VCi$_3$Te$_4$ & 991652 & 1.4492 & 0.5130 & 0.6041 & 24.4210 & $-$3.4238 & 2.0000 & 39.7500 & 4.5000 & 93.9974 \\
KNa$_2$CuO$_2$ & 545359 & 4.3044 & 1.6193 & 3.1185 & 16.5305 & $-$3.3477 & 1.3333 & 14.3333 & 3.0000 & 30.1040 \\
KLiSO$_4$ & 6179 & 8.9145 & 5.5435 & 9.1449 & 14.2480 & $-$4.6714 & 1.7143 & 10.0000 & 2.4286 & 20.3106 \\
Ba$_2$LaSbO$_6$ & 551269 & 6.7642 & 4.1321 & 5.9146 & 16.0099 & $-$5.6548 & 2.4000 & 26.8000 & 3.5000 & 63.1304 \\
SrWO$_4$ & 19163 & 7.1743 & 4.3956 & 5.3508 & 14.4879 & $-$6.6017 & 2.6667 & 24.0000 & 3.1667 & 55.9117 \\
Y$_2$CN$_2$O$_2$ & 546864 & 5.9604 & 4.0829 & 5.8054 & 13.6567 & $-$7.0505 & 2.8571 & 16.2857 & 2.8571 & 35.6893 \\
LiCaAlF$_6$ & 6134 & 12.7564 & 7.9256 & 10.9585 & 11.5957 & $-$5.2724 & 1.3333 & 10.0000 & 2.3333 & 20.8952 \\ 
YScO$_3$ & 768479 & 6.0666 & 3.8188 & 5.1118 & 13.3224 & $-$7.3538 & 2.4000 & 16.8000 & 3.0000 & 36.3724 \\ 
NaZrCu$_3$Se$_4$ & 1180089 & 2.6075 & 1.3386 & 2.1191 & 23.0723 & $-$3.8742 & 1.7778 & 30.4444 & 4.0000 & 68.9658 \\ 
Li$_2$MgCdP$_2$ & 1222661 & 2.2546 & 1.3722 & 2.2428 & 18.0372 & $-$2.5509 & 2.0000 & 16.0000 & 3.0000 & 35.4473 \\ 
SrTaF$_7$ & 555119 & 10.1503 & 5.5729 & 7.4006 & 13.0156 & $-$5.6906 & 1.5556 & 19.3333 & 2.7778 & 44.6171 \\ 
CaMg$_6$ZnO$_8$ & 1032894 & 6.3302 & 3.3602 & 5.2130 & 10.0093 & $-$4.9620 & 2.0000 & 11.6250 & 2.6250 & 23.7061 \\ 
VBO$_4$ & 754594 & 5.7642 & 2.7920 & 3.2228 & 14.7864 & $-$6.5428 & 2.6667 & 10.0000 & 2.3333 & 20.9587 \\
Ca$_2$ZrTiSi$_2$O$_{10}$ & 1227799 & 6.2472 & 3.2485 & 4.0439 & 12.0577 & $-$6.7712 & 2.5000 & 13.1250 & 2.6875 & 27.2144 \\ 
SrZnAsF & 1080135 & 1.9016 & 0.5867 & 1.8223 & 18.7051 & $-$3.4577 & 2.0000 & 27.5000 & 3.7500 & 61.7325 \\
\bottomrule
  \end{tabular*} 
\footnotetext[1]{From Materials Project Database. [Jain $et$ $al$., Appl. Mater. Phys. Lett. 1, 011002 (2013)]}
\footnotetext[2]{In the unit of eV.}
\footnotetext[3]{In the unit of \AA$^3$/atom.}
\footnotetext[4]{In the unit of eV/atom.}
\end{minipage}
\end{center}
\end{table}
\clearpage
\newgeometry{left=1cm,right=1cm,top=1cm,bottom=1.5cm}

\begin{table}[tp]
  \ContinuedFloat
  \begin{center}
  \begin{minipage}{\textwidth}
  \caption{
  (Continued.)
  }   
  \begin{tabular*}{0.87\linewidth}{lcccccccccc}
\toprule
Material & $\bar{r}$ & $\bar{\chi}$ & $\bar{I}$ & $\sigma$($\mid$$n$$\mid$) & $\sigma(Z)$ & $\sigma(p)$ & $\sigma(m)$ & $\sigma(r)$ & $\sigma(\chi)$ & $\sigma(I)$\\
\midrule%
Ba$_2$MgB$_2$O$_6$ & 1.8227 & 2.5282 & 10.5795 & 0.3857 & 18.6809 & 1.5266 & 47.0025 & 0.4314 & 1.0581 & 3.4646 \\
K$_2$LiAlF$_6$ & 1.7980 & 2.8110 & 12.4596 & 0.6000 & 4.6861 & 0.8062 & 9.4045 & 0.4960 & 1.4465 & 6.0951 \\
KAlSiO$_4$ & 1.8243 & 2.5843 & 10.4215 & 0.8806 & 4.0153 & 0.7284 & 8.4151 & 0.4315 & 1.0323 & 3.8299 \\
KLi$_6$BiO$_6$ & 1.7757 & 2.0886 & 8.9775 & 1.0302 & 20.1282 & 1.1157 & 50.9776 & 0.3207 & 1.1945 & 4.0596 \\
CaTiO$_3$ & 1.7960 & 2.5720 & 10.7591 & 0.8000 & 6.4000 & 0.9798 & 13.9271 & 0.3439 & 1.0767 & 3.5088 \\
Ag$_2$GePbS$_4$ & 1.9438 & 2.2488 & 8.9886 & 0.8660 & 22.1980 & 1.1180 & 58.0362 & 0.1464 & 0.3355 & 1.3770 \\
NaAlS$_2$O$_8$ & 1.6558 & 2.9350 & 11.7325 & 0.4082 & 3.0822 & 0.4714 & 6.3013 & 0.2237 & 0.8200 & 3.0098 \\
AgAlS$_2$ & 1.8875 & 2.1750 & 8.5705 & 0.7071 & 13.9104 & 0.8660 & 33.6213 & 0.1295 & 0.4205 & 1.8758 \\
NbCu$_3$Se$_4$ & 1.9575 & 2.1875 & 8.6185 & 1.2247 & 3.8079 & 0.3307 & 9.8554 & 0.0886 & 0.3740 & 1.1720 \\
Ba$_2$SrWO$_6$ & 1.9150 & 2.5070 & 10.5691 & 1.2000 & 24.8548 & 1.8574 & 62.8484 & 0.5007 & 1.1631 & 3.7981 \\
RbSrCO$_3$F & 1.8929 & 2.6600 & 11.3442 & 0.9258 & 13.4453 & 1.3553 & 32.0184 & 0.5711 & 1.1886 & 4.4037 \\
CaYAlO$_4$ & 1.7929 & 2.5129 & 10.3984 & 0.4518 & 10.6962 & 1.1249 & 24.9296 & 0.3475 & 1.0832 & 3.7183 \\
Sr$_2$ScGaO$_5$ & 1.8444 & 2.4744 & 10.2268 & 0.4157 & 12.7976 & 1.2862 & 30.3908 & 0.4019 & 1.1052 & 3.7989 \\
Sr$_2$GaSbO$_6$ & 1.8030 & 2.6400 & 10.7706 & 0.9165 & 16.0947 & 1.3748 & 38.9417 & 0.3867 & 1.0290 & 3.5723 \\
CdCu$_2$GeS$_4$ & 1.9263 & 2.2275 & 9.2233 & 0.8660 & 10.8022 & 0.6960 & 27.0556 & 0.1433 & 0.3619 & 1.1966 \\
Sr$_2$NbInO$_6$ & 1.8210 & 2.5920 & 10.5644 & 0.9165 & 16.6565 & 1.4697 & 39.7054 & 0.3974 & 1.0654 & 3.7508 \\
RbNaTiO$_3$ & 1.9950 & 2.2683 & 9.4998 & 1.0000 & 10.7497 & 1.1547 & 25.6086 & 0.5533 & 1.1929 & 4.1906 \\
Li$_2$CdGeO$_4$ & 1.7513 & 2.4275 & 10.2686 & 0.8660 & 15.2541 & 1.1110 & 36.0241 & 0.2588 & 1.0612 & 3.5299 \\
CeSiO$_4$ & 1.7667 & 2.7967 & 11.3605 & 0.9428 & 18.3182 & 1.4625 & 45.5683 & 0.3609 & 0.9373 & 3.2807 \\
CdAg$_2$I$_4$ & 2.0457 & 2.3129 & 9.4216 & 0.3499 & 2.8212 & 0.0000 & 8.7836 & 0.0789 & 0.4076 & 1.2669 \\
Rb$_2$Li$_2$GeO$_4$ & 1.9878 & 2.1522 & 9.0566 & 0.9162 & 13.8724 & 1.2862 & 32.5143 & 0.5887 & 1.1996 & 4.2038 \\
SrTiO$_3$ & 1.8320 & 2.5620 & 10.6755 & 0.8000 & 11.9063 & 1.2649 & 28.2960 & 0.4006 & 1.0914 & 3.6218 \\
ZnCu$_2$SiTe$_4$ & 2.0338 & 1.9688 & 8.6296 & 0.8660 & 14.0601 & 0.6960 & 37.8793 & 0.0482 & 0.1519 & 0.6146 \\
SrAgTeF & 2.0325 & 2.2400 & 9.9259 & 0.5000 & 16.6508 & 1.2990 & 40.9260 & 0.3649 & 1.0963 & 4.4851 \\
Li$_2$ZnSnS$_4$ & 1.8775 & 1.9863 & 8.6202 & 0.8660 & 14.2719 & 0.9270 & 34.0281 & 0.1293 & 0.6661 & 2.0971 \\
VCi$_3$Te$_4$ & 2.0237 & 1.9663 & 8.2455 & 1.2247 & 12.3870 & 0.5000 & 33.8235 & 0.0495 & 0.1572 & 0.8210 \\
KNa$_2$CuO$_2$ & 2.0483 & 1.9100 & 8.2636 & 0.4714 & 7.5203 & 0.8165 & 16.8242 & 0.4391 & 1.1395 & 3.9273 \\
KLiSO$_4$ & 1.7786 & 2.5914 & 10.6521 & 0.4518 & 5.0990 & 0.7284 & 10.2983 & 0.4163 & 1.1094 & 3.8318 \\
Ba$_2$LaSbO$_6$ & 1.8970 & 2.5570 & 10.6317 & 0.9165 & 23.0729 & 1.8574 & 57.8923 & 0.4888 & 1.1232 & 3.7671 \\
SrWO$_4$ & 1.7917 & 2.7350 & 11.3386 & 1.4907 & 24.8998 & 1.6750 & 62.9092 & 0.3945 & 1.0203 & 3.2840 \\
Y$_2$CN$_2$O$_2$ & 1.7829 & 2.5643 & 11.4285 & 0.6389 & 14.3797 & 1.3553 & 33.6809 & 0.3445 & 0.8945 & 3.4480 \\
LiCaAlF$_6$  & 1.6433 & 3.0522 & 13.5586 & 0.6667 & 4.2687 & 0.6667 & 8.2960 & 0.2778 & 1.3229 & 5.4678 \\
YScO$_3$ & 1.8060 & 2.5800 & 10.7266 & 0.4899 & 12.1885 & 1.2649 & 28.5607 & 0.3544 & 1.0542 & 3.5430 \\
NaZrCu$_3$Se$_4$ & 1.9978 & 2.0178 & 8.2180 & 0.9162 & 7.6465 & 0.4714 & 18.5498 & 0.1377 & 0.5608 & 1.5681 \\
Li$_2$MgCdP$_2$  & 1.8583 & 1.5567 & 8.0661 & 0.8165 & 15.1658 & 1.0000 & 35.8351 & 0.1470 & 0.5076 & 2.1237 \\
SrTaF$_7$ & 1.6667 & 3.3678 & 15.0227 & 1.2571 & 21.0185 & 1.4741 & 52.7360 & 0.3734 & 1.1527 & 4.5114 \\
CaMg$_6$ZnO$_8$ & 1.6788 & 2.3769 & 10.6456 & 0.0000 & 5.6665 & 0.6960 & 12.4570 & 0.2130 & 1.0693 & 3.0287 \\
VBO4 & 1.6783 & 2.9050 & 11.5861 & 1.1055 & 5.9161 & 0.7454 & 13.5417 & 0.2281 & 0.7658 & 2.9084 \\
Ca$_2$ZrTiSi$_2$O$_{10}$ & 1.7725 & 2.6919 & 11.1358 & 0.8660 & 8.5138 & 0.9823 & 19.5486 & 0.3310 & 0.9924 & 3.2472 \\
SrZnAsF & 1.9550 & 2.1900 & 10.5751 & 0.7071 & 11.0567 & 1.0897 & 25.9024 & 0.3659 & 1.1218 & 4.2638 \\
\bottomrule
  \end{tabular*} 
\end{minipage}
\end{center}
\end{table}

\clearpage
\newgeometry{left=1cm,right=1cm,top=1cm,bottom=1.5cm}
\begin{table}[tp]
  \begin{center}
  \begin{minipage}{\textwidth}
  \caption{
  Sequential order of elimination of seven features from the pristine \textit{18-feature set} prior to XML analyses.
  }
  \label{tab:feature_elimination}
  \begingroup
  \renewcommand{\arraystretch}{1.0}
 \resizebox{\linewidth}{!}{%
  \begin{tabular}{c c c c c c c c c c c c c c}
    \toprule
    \multicolumn{1}{c}{\# Features} & \multicolumn{1}{c}{$r_p$} & \multicolumn{1}{c}{$r_s$} &
    \multicolumn{4}{c}{Retained feature} &
    \multicolumn{4}{c}{Eliminated feature} &
    \multicolumn{1}{c}{$p$-value (pairwise} &  \multicolumn{1}{c}{$p$-value ($t$-test} & Max.\\
    \cmidrule(lr){4-7}\cmidrule(lr){8-11}
    at test & & & Feature & RMSE (eV) $^($\footnotemark[1]$^)$ & $r_p$ vs.\ $E_\mathrm{g}^\mathrm{GW}$ & $r_s$ vs.\ $E_\mathrm{g}^\mathrm{GW}$ & Feature & RMSE (eV) $^($\footnotemark[1]$^)$ & $r_p$ vs.\ $E_\mathrm{g}^\mathrm{GW}$ & $r_s$ vs.\ $E_\mathrm{g}^\mathrm{GW}$ & $t$-test) $^($\footnotemark[2]\footnotemark[3]$^)$ & vs. pristine) $^($\footnotemark[4]$^)$ & VIF $^($\footnotemark[5]$^)$ \\
    \midrule
    18 & 1.00   & 1.00   & $\overline{Z}$ & 0.252 $\pm$ 0.024 & $-$0.52 & $-$0.55 & $\overline{m}$ & 0.252 $\pm$ 0.025 & $-$0.51 & $-$0.53 & 0.761 $^($\footnotemark[3]$^)$ & 0.091 & 448  \\
    17 & 1.00 & 0.99 & $\sigma(m)$ & 0.245 $\pm$ 0.025 & $-$0.37 & $-$0.37 & $\sigma(Z)$ & 0.246 $\pm$ 0.025 & $-$0.35 & $-$0.34 & 0.129 $^($\footnotemark[3]$^)$ & 0.515 & 78.1 \\
    16 & 0.97 & 0.97 & $E_\mathrm{g}^\mathrm{PBE}$ & 0.387 $\pm$ 0.042 &  0.98   &  0.98   & $E_\mathrm{g}^\mathrm{mBJ}$ & 0.255 $\pm$ 0.033 & 0.98 & 0.98 & 1.9$\times$10$^{-9}$ $^($\footnotemark[2]$^)$ & 0.096 & 74.7 \\
    15 & 0.97 & 0.96 & $\overline{Z}$ & 0.262 $\pm$ 0.036 & $-$0.52 & $-$0.55 & $\overline{p}$ & 0.263 $\pm$ 0.037 & $-$0.52 & $-$0.51 & 0.040 $^($\footnotemark[3]$^)$ & 0.020 & 17.2  \\
    14 & 0.95 & 0.94 & $\overline{\chi}$ & 0.268 $\pm$ 0.026 &  0.57   &  0.48   & $\overline{I}$      & 0.251 $\pm$ 0.027 &  0.60   &  0.49   & 0.001 $^($\footnotemark[2]$^)$ & 0.195 & 16.8 \\
    13 & 0.88 & 0.87 & $\sigma(\chi)$ & 0.271 $\pm$ 0.028 &  0.80   &  0.81   & $\sigma(I)$  & 0.258 $\pm$ 0.028 &  0.74   &  0.71   & 0.013 $^($\footnotemark[3]$^)$ & 0.020 & 8.1 \\
    12 & 0.77 & 0.80 & $\overline{Z}$ & 0.252 $\pm$ 0.027 & $-$0.52 & $-$0.55 & $V$   & 0.255 $\pm$ 0.025 & $-$0.22 & $-$0.25 & 0.514 $^($\footnotemark[3]$^)$ & 0.063 & 7.9 \\
    \midrule
    11 & 0.73 & 0.74 & $E_\mathrm{g}^\mathrm{PBE}$  & 0.960 $\pm$ 0.131 &  0.98   &  0.98   & $\sigma(\chi)$   & 0.289 $\pm$ 0.022 $^($\footnotemark[4]$^)$ &  0.80   &  0.81   & 2.4$\times$10$^{-15}$ & 1.1$\times$10$^{-7}$ & 4.9 \\
    \bottomrule
  \end{tabular}} 
  \endgroup
      \footnotetext[1]{ The RMSE obtained using the predictive model when this feature is removed. 
      $\pm$ indicates one standard deviation across the predictive models with 20 different data selections.}
      \footnotetext[2]{ The elimination is based on a statistically significant increase (99\% confidence level) in the prediction error ($p$-value $<$ 0.01).}
   \footnotetext[3]{ The elimination is based on lower correlation with the prediction objective ($E_\mathrm{g}^\mathrm{GW}$) as no significant error difference is observed ($p$-value $>$ 0.01). }
    \footnotetext[4]{ 
    At each elimination step, the prediction error of the reduced feature set is compared with that of the pristine model using a paired $t$-test.
    If the removal of a feature does not result in a statistically significant increase in prediction error ($p$-value $>$ 0.01), the elimination process proceeds to the next step.
    If a statistically significant increase is observed ($p$-value $<$ 0.01), the elimination process is terminated.
    Consequently, the feature set is reduced to 11 features.}
    \footnotetext[5] {The maximum variance inflation factor (VIF) at each elimination step is additionally reported to illustrate the progressive reduction of multicollinearity.
    The reported value corresponds to the maximum VIF after the removal of one feature at each stage. For the original 18-feature set, the maximum VIF is 8135.}

  \end{minipage}
  \end{center}
\end{table}
\restoregeometry
\clearpage
\clearpage
\begin{table}[ht]
  \begin{center}
  \begin{minipage}{\textwidth}
  \caption{
  Prediction errors for test dataset using SVR models with different numbers of features.
  $\pm$ indicates one standard deviation across the predictive models with 20 different data selections.
  }
  \label{tab:svr_performance_raw}
  \begingroup
  \renewcommand{\arraystretch}{1.1}
  \resizebox{\linewidth}{!}{%
\begin{tabular}{l c c c c c c}
\toprule
Number of & \multicolumn{3}{c}{In-domain} & \multicolumn{3}{c}{OOD}\\
\cmidrule(lr){2-4}\cmidrule(lr){5-7}
features &  RMSE (eV) & MAE (eV) & R$^2$ & RMSE (eV) & MAE (eV) & R$^2$ \\
\midrule
18 &  0.247 $\pm$ 0.019 & 0.184 $\pm$ 0.013 & 0.993 $\pm$ 0.001 & 0.460 $\pm$ 0.053 & 0.376 $\pm$ 0.042 & 0.980 $\pm$ 0.004 \\
11 &  0.255 $\pm$ 0.025 & 0.188 $\pm$ 0.018 & 0.993 $\pm$ 0.001 & 0.457 $\pm$ 0.037 & 0.373 $\pm$ 0.026 & 0.983 $\pm$ 0.003 \\
10 &  0.254 $\pm$ 0.023 & 0.187 $\pm$ 0.015 & 0.993 $\pm$ 0.001 & 0.473 $\pm$ 0.069 & 0.382 $\pm$ 0.049 & 0.981 $\pm$ 0.007 \\
9 &  0.253 $\pm$ 0.021 & 0.186 $\pm$ 0.014 & 0.993 $\pm$ 0.001 & 0.483 $\pm$ 0.103 & 0.386 $\pm$ 0.065 & 0.979 $\pm$ 0.010 \\
8 &  0.266 $\pm$ 0.031 & 0.195 $\pm$ 0.017 & 0.992 $\pm$ 0.002 & 0.398 $\pm$ 0.053 & 0.316 $\pm$ 0.037 & 0.982 $\pm$ 0.004 \\
7 &  0.260 $\pm$ 0.025 & 0.189 $\pm$ 0.016 & 0.993 $\pm$ 0.002 & 0.335 $\pm$ 0.019 & 0.279 $\pm$ 0.014 & 0.987 $\pm$ 0.001 \\
6 & 0.259 $\pm$ 0.026 & 0.189 $\pm$ 0.016 & 0.993 $\pm$ 0.002 & 0.354 $\pm$ 0.039 & 0.292 $\pm$ 0.032 & 0.985 $\pm$ 0.003 \\
5 &  0.254 $\pm$ 0.020 & 0.191 $\pm$ 0.014 & 0.993 $\pm$ 0.001 & 0.348 $\pm$ 0.014 & 0.287 $\pm$ 0.012 & 0.986 $\pm$ 0.000 \\
4 &  0.289 $\pm$ 0.021 & 0.217 $\pm$ 0.015 & 0.991 $\pm$ 0.002 & 0.427 $\pm$ 0.027 & 0.355 $\pm$ 0.021 & 0.982 $\pm$ 0.001 \\
3 &  0.312 $\pm$ 0.032 & 0.234 $\pm$ 0.023 & 0.989 $\pm$ 0.003 & 0.486 $\pm$ 0.021 & 0.393 $\pm$ 0.018 & 0.977 $\pm$ 0.002 \\
2 &  0.453 $\pm$ 0.045 & 0.313 $\pm$ 0.033 & 0.978 $\pm$ 0.004 & 0.483 $\pm$ 0.019 & 0.388 $\pm$ 0.020 & 0.969 $\pm$ 0.002 \\
\bottomrule
  \end{tabular}}
  \endgroup
  \end{minipage}
  \end{center}
\end{table}
\clearpage

\end{document}